\documentclass[superscriptaddress,onecolumn,pre]{revtex4}

\usepackage{amsmath}
\usepackage{graphicx,color}

\newcommand{\be}{\begin{equation}}
\newcommand{\ee}{\end{equation}}
\newcommand{\bea}{\begin{eqnarray}}
\newcommand{\eea}{\end{eqnarray}}

\begin{document}
\sloppy

%-title page-%

\title{Kinetic theory of long-range interacting systems with angle-action \\
variables and collective effects}

\author{Pierre-Henri Chavanis}
%\email{chavanis@irsamc.ups-tlse.fr}
\affiliation{Laboratoire de Physique Th\'eorique (IRSAMC), CNRS and UPS, Universit\'e de Toulouse, F-31062 Toulouse, France}

\begin{abstract}

We develop a kinetic theory of systems with long-range interactions
taking collective effects and spatial inhomogeneity into
account. Starting from the Klimontovich equation and using a
quasilinear approximation, we derive a Lenard-Balescu-type kinetic
equation written in angle-action variables. We confirm the result
obtained by Heyvaerts [Mon. Not. R. Astron. Soc. {\bf 407}, 355
(2010)] who started from the Liouville equation and used the BBGKY
hierarchy truncated at the level of the two-body distribution
function. When collective effects are neglected, we recover the
Landau-type kinetic equation obtained in our previous papers
[P.H. Chavanis, Physica A {\bf 377}, 469 (2007); J. Stat. Mech.,
P05019 (2010)]. We also consider the relaxation of a test particle in
a bath of field particles. Its stochastic motion is described by a
Fokker-Planck equation written in angle-action variables. We determine
the diffusion tensor and the friction force by explicitly calculating
the first and second order moments of the increment of action of the
test particle from its equations of motion, taking collective effects
into account. This generalizes the expressions obtained in our
previous works. We discuss the scaling with $N$ of the relaxation time
for the system as a whole and for a test particle in a bath.

\end{abstract}

\maketitle

\section{Introduction}
\label{sec_introduction}

Systems with long-range interactions are numerous in nature. Some
examples include self-gravitating systems, two-dimensional vortices,
non-neutral plasmas and the Hamiltonian Mean Field (HMF) model
\cite{houches}.
The equilibrium statistical mechanics of systems with long-range
interactions is now relatively well understood \cite{cdr}. One
important feature of these systems is that they may exhibit ensemble
inequivalence and negative specific heats first observed in
astrophysics (see \cite{katz,paddy,ijmpb} for reviews).  The
out-of-equilibrium properties of these systems are also very
interesting \cite{cdr}. In particular, long-range interacting systems may be
stuck in non-Boltzmannian quasi stationary states (QSS) that persist
for a very long time before finally reaching the Boltzmann
distribution. These QSSs are stable steady states of the Vlasov
equation which governs the ``collisionless'' evolution of the
system. In order to understand the different timescales involved in
the dynamics, it is important to develop a kinetic theory of systems
with long-range interactions. Different kinetic theories have been proposed
over the years.

Boltzmann \cite{boltzmann} developed a kinetic theory for dilute gases.
In that case, the particles do not interact except during hard binary
collisions. Boltzmann derived a kinetic equation from which
he proved that entropy increases ($H$-theorem) and
that the system ultimately relaxes towards the Maxwellian distribution of
statistical equilibrium.

Landau \cite{landau} derived a kinetic equation for a neutral
Coulombian plasma in which the charges interact via a $1/r$ potential.
He started from the Boltzmann equation and considered a weak
deflection approximation. Since the dynamical evolution of the system
is dominated by weak binary collisions, it is possible to expand the
Boltzmann equation in terms of a small deflection parameter $\Delta\ll
1$ and make a linear trajectory approximation. However, this treatment
yields a logarithmic divergence at small and large impact parameters.
The divergence at small scales is due to the neglect of strong
collisions, resulting in large deflections. It can be cured by
introducing a cut-off at the Landau length, corresponding to the
impact parameter leading to a deflection at $90^{o}$. On the other
hand, in a neutral plasma, the potential is screened on a distance
corresponding to the Debye length \cite{dh}. Phenomenologically, the Debye
length provides an upper cut-off. Later on, Lenard \cite{lenard} and
Balescu
\cite{balescu} developed a more precise  kinetic
theory that takes collective effects into account. This amounts to
replacing the bare potential of interaction $\hat{u}({\bf k})$ by a
``dressed'' potential of interaction $\hat{u}({\bf k})/|\epsilon({\bf
k},{\bf k}\cdot{\bf v})|$ where $\epsilon({\bf k},{\bf k}\cdot{\bf
v})$ is the dielectric function. Physically, this means that the
particles are ``dressed'' by their polarization cloud. The original Landau
equation, which ignores collective effects, is recovered from the
Lenard-Balescu equation by setting $|\epsilon({\bf k},{\bf k}\cdot{\bf
v})|=1$. However, when collective effects are taken into account, it
is found that the logarithmic divergence that appears at large scales
in the Landau equation is removed and that the Debye length is indeed
the natural upper lengthscale to consider.  At about the same time,
Hubbard \cite{hubbard} developed a test particle approach and derived
a Fokker-Planck equation describing the relaxation of a test particle
in a bath of field particles. He calculated the  diffusion and friction
coefficients by evaluating the first and second moments of the
velocity deflection and took collective effects into account. The
Fokker-Planck equation of Hubbard is recovered from the Lenard-Balescu
equation in the ``bath'' approximation.  These kinetic
theories lead to a relaxation time towards the Maxwell-Boltzmann distribution
of the form $t_{R}\sim (\Lambda/\ln
\Lambda) t_D$,  where $t_D$ is the dynamical time (the inverse of the plasma pulsation) and $\Lambda$ the plasma
parameter (the number of charges in the Debye sphere).

Chandrasekhar \cite{chandra,chandra1,chandra2} developed a kinetic
theory of stellar systems in order to determine the timescale of
collisional relaxation and the rate of escape of stars from globular
clusters. To simplify the kinetic theory, he considered an infinite
homogeneous system. He started from the general Fokker-Planck equation
and determined the diffusion coefficient and the friction force (first
and second moments of the velocity increment) by considering the mean
effect of a succession of two-body encounters. Since his approach can
take large deflections into account, there is no divergence at small
impact parameters and the gravitational analogue of the Landau length
naturally appears in the treatment of Chandrasekhar.  However, his
approach leads to a logarithmic divergence at large scales that is
more difficult to remove in stellar dynamics than in plasma physics
because of the absence of Debye shielding for the gravitational
force. In a series of papers, Chandrasekhar and von Neumann \cite{cn}
developed a completely stochastic formalism of gravitational
fluctuations and showed that the fluctuations of the gravitational
force are given by the Holtzmark distribution (a particular L\'evy
law) in which the nearest neighbor plays a prevalent role. From these
results, they argued that the logarithmic divergence has to be cut-off
at the interparticle distance. However, since the interparticle
distance is smaller than the Debye length, the same arguments should
also apply in plasma physics, which is not the case. Therefore, the
conclusions of Chandrasekhar and von Neumann are usually taken with
circumspection. In particular, Cohen {\it et al.} \cite{cohen} argue
that the logarithmic divergence should be cut-off at the Jeans length
\cite{jeansbook}, giving an estimate of the system's size, since the
Jeans length is the gravitational analogue of the Debye
length. Indeed, while in neutral plasmas the effective interaction
distance is limited to the Debye length, in a self-gravitating system,
the distance between interacting particles is only limited by the
systems's size. These kinetic theories lead to a relaxation time of
the form $t_{R}\sim (N/\ln N) t_D$ where $t_D$ is the dynamical time
and $N$ the number of stars in the system.  Chandrasekhar
\cite{nice} also developed a Brownian theory of stellar dynamics and
showed that, from a qualitative point of view, the results of kinetic
theory can be understood very simply in that framework. In particular,
he showed that a dynamical friction is necessary to reproduce the
Maxwell-Boltzmann distribution of statistical equilibrium and that the
coefficients of friction and diffusion are related to each other by an
Einstein relation (fluctuation-dissipation theorem). This relation is
confirmed by his more precise kinetic theory based on two-body
encounters.  It is important to emphasize, however, that Chandrasekhar
did not derive the kinetic equation for the evolution of the system as
a whole. Indeed, he considered the Brownian motion of a test star in a
fixed distribution of field stars (bath) and derived the corresponding
Fokker-Planck equation \footnote{Indeed, Chandrasekhar
\cite{chandra1,chandra2} models the evolution of
globular clusters by the Kramers  \cite{kramers} equation which has a fixed
temperature (canonical description) while a more relevant kinetic
equation would be the Landau \cite{landau} equation which conserves energy
(microcanonical description).}. This equation has been used 
by Chandrasekhar \cite{chandra2}, Spitzer and H\"arm \cite{spitzer},
Michie \cite{michie}, King \cite{king}, and Lemou and Chavanis
\cite{lcstar} to study the evaporation of stars from globular
clusters. King \cite{kingL} noted that, if we were to describe the
dynamical evolution of the cluster as a whole, the distribution of the
field particles must evolve in time in a self-consistent manner so
that the kinetic equation must be an integrodifferential equation. The
kinetic equation obtained by King is equivalent to the Landau
equation. There is, however, an important difference between stellar
dynamics and plasma physics. Neutral plasmas are spatially homogeneous
due to Debye shielding.  By contrast, stellar systems are spatially
inhomogeneous. The above-mentioned kinetic theories developed for an
infinite homogeneous system can be applied to an inhomogeneous system
only if we make a {\it local approximation}. In that case, the
collision term is calculated as if the system were spatially
homogeneous or as if the collisions could be treated as local. Then,
the effect of spatial inhomogeneity is only retained in the advective
(Vlasov) term which describes the evolution of the system due to
mean-field effects \cite{jeans,vlasov}. This leads to the
Vlasov-Landau-Poisson system which is the standard kinetic equation of
stellar dynamics. To our knowledge, this equation has been first
written and studied by H\'enon
\cite{henon}. H\'enon also exploited the timescale separation between
the dynamical time $t_D$ and the relaxation time $t_{R}\gg t_D$ to
derive a simplified kinetic equation for $f(\epsilon,t)$, where
$\epsilon=v^2/2+\Phi({\bf r},t)$ is the individual energy of a star by
unit of mass, called the orbit-averaged Fokker-Planck equation. In
this approach, the distribution function $f({\bf r},{\bf v},t)$,
averaged over a short timescale, is a steady state of the Vlasov
equation of the form $f(\epsilon,t)$ which slowly evolves in time, on
a long timescale, due to the development of ``collisions''
(i.e. correlations caused by finite $N$ effects or graininess).  Cohn
\cite{cohn} numerically solved the orbit-averaged Fokker-Planck
equation to describe the collisional evolution of star clusters
\footnote{The orbit-averaged approximation has been recently
criticized by Lancellotti and Kiessling \cite{lk}.}.  His treatment
accounts both for the escape of high energy stars put forward by
Spitzer \cite{spitzerevap}, and for the phenomenon of core collapse
resulting from the gravothermal catastrophe discovered by Antonov
\cite{antonov} and Lynden-Bell and Wood \cite{lbw} on the basis of
statistical mechanics.  The local approximation, which is a crucial
step in the kinetic theory, is supported by the stochastic approach of
Chandrasekhar and von Neumann
\cite{cn} showing the preponderance of the nearest
neighbor. However, this remains a simplifying assumption which is not
easily controllable. In particular, as we have already indicated, the
local approximation leads to a logarithmic divergence at large scales
that is difficult to remove. This divergence would not have occurred
if full account of spatial inhomogeneity had been given since the
start. The effect of spatial inhomogeneity was investigated by Severne
and Haggerty \cite{severne}, Parisot and Severne \cite{ps}, Kandrup
\cite{kandrup1}, and Chavanis \cite{paper3,paper4}. In particular,
Kandrup
\cite{kandrup1} derived a generalized Landau equation from the Liouville equation by using
projection operator technics. Recently, Chavanis \cite{paper3,paper4}
obtained this equation in a simpler manner from the BBGKY hierarchy or
from a quasilinear theory, as an expansion in $1/N$ in a proper
thermodynamic limit.  This generalized Landau equation is interesting
because it takes into account effects of spatial inhomogeneity and
memory which were neglected in previous approaches. Since the finite
extension of the system is properly accounted for, there is no
divergence at large scales \footnote{There remains, however, a
logarithmic divergence at small scales due to the neglect of strong
collisions.}. Furthermore, this approach clearly shows which
approximations are needed in order to recover the traditional Landau
equation. Unfortunately, the generalized Landau equation remains
extremely complicated for practical applications. In addition, this
equation is still approximate as it neglects ``collective effects''
and considers binary collisions between naked particles. As in any
weakly coupled system, the particles engaged in collisions are dressed
by the polarization clouds caused by their own influence on other
particles. Collisions between dressed particles have quantitatively
different outcomes than collisions between naked ones. In the case of
plasmas, collective effects are responsible for Debye shielding.  For
self-gravitating systems, they lead to ``anti-shielding'' and are more
difficult to analyze \footnote{In a plasma, since the Coulomb force
between electrons is repulsive, each particle of the plasma tends to
attract to it particles of opposite charge and to repel particles of
like charge, thereby creating a kind of ``cloud'' of opposite charge
which screens the interaction at the scale of the Debye length. In the
gravitational case, since the Newton force is attractive, the
situation is considerably different. The test star draws neighboring
stars into its vicinity and these add their gravitational force to
that of the test star itself. The ``bare'' gravitatonal force of the
test star is thus augmented rather than shielded. The polarization
acts to increase the effecive gravitational mass of a test
star.}. Some authors like Thorne \cite{thorne}, Miller
\cite{rhmiller}, Gilbert \cite{gilbert1,gilbert2}, Lerche
\cite{lerche}, and Weinberg \cite{weinberglb} attempted to take both
spatial inhomogeneity and collective effects into account. However,
they obtained very complicated kinetic equations that have not found
application until now. They managed, however, to show that collective
effects are equivalent to increasing the effective mass of the
particles, hence diminishing the relaxation time. Since, on the other
hand, the effect of spatial inhomogeneity is to increase the
relaxation time \cite{ps}, the two effects act in opposite direction
and may balance each other.

The same difficulties (and additional ones) occur for other systems with long-range
interactions \cite{cdr}. For example, the kinetic theory of the HMF model \footnote{This model \cite{ar} is directly inspired by astrophysics. It was proposed by Pichon \cite{pichonphd} as a simplified model to describe the formation of bars in disk galaxies (for an historic of the HMF model, see \cite{cvb}).} is very complicated. In the homogeneous phase (at supercritical energies) the Lenard-Balescu collision term vanishes \cite{bd,cvb}, like in the case of homogeneous 1D plasmas \cite{feix,kp}, so the kinetic theory must take  higher order correlations (three-body, four-body...) into account. In the inhomogeneous phase (at subcritical energies), we have to face the problems of spatial delocalization discussed above. The kinetic theory of 2D point vortices \cite{dubin,pre,pvnew} presents very similar features. Finally, 1D stellar systems are always spatially inhomogeneous and it is not possible to make a local approximation, contrary to the 3D case.

An important problem in the kinetic theory of systems with long-range
interactions is therefore to derive a kinetic equation that takes both
spatial inhomogeneity and collective effects into account. A standard
tool to deal with spatially inhomogeneous systems is to use
angle-action variables. This has been used by many authors in
astrophysics in order to solve dynamical stability problems
\cite{kalnajsdisp,goodman88,weinbergdisp,pichoncannon,valageas1} or
to compute diffusion and friction coefficients in a Fokker-Planck
approach \cite{lbk,tw,bl,wei,nt,w,pa,valageas2}. In an earlier paper
\cite{angleaction}, starting from the Klimontovich equation and using
a quasilinear approximation, we attempted to derive a
Lenard-Balescu-type equation with angle-action variables, taking
collective effects into account. However, in the course of the
derivation, we made a factorization assumption that was not
justified. We had the intuition that this problem might be solved by
expanding the density-potential functions on a ``good'' basis but we
did not succeed in developing an adequate formalism. However, we
argued that the final kinetic equation that we obtained (see Eq. (37)
of \cite{angleaction}) is correct when collective effects are
neglected. In a subsequent paper \cite{kindetail}, we derived this
kinetic equation in a more satisfactory manner without {\it ad hoc}
assumption. Since collective effects are neglected, this kinetic
equation can be viewed as a generalization of the Landau equation,
written in angle-action variables, applying to spatially inhomogeneous
systems with long-range interactions \footnote{We considered 1D
systems to simplify the notations but the formal generalization in $d$
dimensions is straightforward.}.

In the meantime, Heyvaerts \cite{heyvaerts} managed to derive a
Lenard-Balescu-type kinetic equation with angle-action variables
starting from the Liouville equation and using the BBGKY hierarchy
truncated at the level of the two-body distribution function
(i.e. neglecting three-body correlations). To calculate the collective
response, he used a biorthogonal basis of pairs of density-potential
functions. This is a relatively classical tool in astrophysics
\cite{btnew}, introduced by Kalnajs
\cite{kalnajs}, that we did not know at the time of our papers
\cite{angleaction,kindetail}. Although the derivation given by
Heyvaerts \cite{heyvaerts} is very satisfactory, we will show in the first part of the
present paper (Sect. \ref{sec_whole}) that the Lenard-Balescu-type kinetic equation can also
be derived from the Klimontovich approach initiated in
\cite{angleaction}. The use of a biorthogonal basis of pairs of
density-potential functions solves the shortcomings of our previous
derivation. This new derivation may be technically simpler than the
one given by Heyvaerts \cite{heyvaerts}. Furthermore, this derivation
is completely parallel to the one exposed in
\cite{pitaevskii,cdr,newpaper} to derive the ordinary Lenard-Balescu
equation and in \cite{dubin,pvnew} to derive a Lenard-Balescu-type
equation appropriate to two-dimensional point vortices. Our approach
provides therefore a unified description of kinetic theories for
systems with long-range interactions when collective effects are
accounted for. This generalizes the kinetic theory presented in
\cite{kindetail} when collective effects are neglected. Our
quasilinear theory has some similarities with the theories developed
by Weinberg \cite{w}, Ma and Bertschinger \cite{mab}, and Pichon and
Aubert \cite{pa} in astrophysics. However, in our case, the source of
noise is due to discreteness (finite $N$) effects internal to the
system while in
\cite{w,mab,pa} it is caused by external sources (perturbations on a
galaxy, cosmological environment on dark matter halos...). In the
second part of the paper (Sect. \ref{sec_stoch}), we consider the
relaxation of a test particle in a bath of field particles. Its
stochastic motion is described by a Fokker-Planck equation with
angle-action variables. We determine the diffusion tensor and the
friction force by explicitly calculating the first and second order
moments of the increment of action of the test particle from its
equations of motion, taking collective effects into account. This
generalizes the expressions obtained in our previous works
\cite{angleaction,kindetail}, neglecting collective effects. We show
the connection between the Fokker-Planck equation describing the
relaxation of a test particle in a bath and the kinetic equation
describing the evolution of the system as a whole. We also discuss the
scaling with $N$ of the relaxation time for the system as a whole and
for a test particle in a bath.

\section{Collisional evolution of the system as a whole}
\label{sec_whole}

\subsection{The Klimontovich equation} \label{sec_klim}

We consider an isolated system of material particles in interaction
moving in a $d$-dimensional space. Their dynamics is described by the
Hamilton equations
\begin{eqnarray}
m\frac{d{\bf r}_i}{dt}=\frac{\partial {\cal H}}{\partial {\bf v}_i},\qquad m \frac{d{\bf v}_i}{dt}=-\frac{\partial {\cal H}}{\partial {\bf r}_i},
\label{pvm1}
\end{eqnarray}
with the Hamiltonian
\begin{eqnarray}
{\cal H}=\sum_{i=1}^{N} \frac{1}{2}mv_i^2+m^2\sum_{i<j}u(|{\bf r}_i-{\bf r}_j|),
\label{pvm2}
\end{eqnarray}
where $u(|{\bf r}-{\bf r}'|)$ is the potential of interaction and $m$ the individual mass of the particles. For an isolated system, the total energy $E={\cal H}$ is conserved. We assume that the potential of interaction decays at large distances like $r^{-\gamma}$ with $\gamma\le d$.  In that case, the potential  is said to be long-ranged \cite{cdr}. Long-range potentials include the gravitational and the Coulombian potentials, corresponding to $\gamma=d-2$.

We introduce the discrete distribution function $f_d({\bf r},{\bf
v},t)=m\sum_i\delta({\bf r}-{\bf r}_i(t))\delta({\bf v}-{\bf v}_i(t))$
and the corresponding potential $\Phi_{d}({\bf r},t)=\int u(|{\bf
r}-{\bf r}'|)f_{d}({\bf r}',{\bf v}',t)d{\bf r}'d{\bf v}'$. Using the
equations of motion (\ref{pvm1}) and (\ref{pvm2}), we find that $f_d({\bf
r},{\bf v},t)$ satisfies the Klimontovich \cite{klimontovich} equation
\begin{equation}
\label{klim2}
\frac{\partial f_{d}}{\partial t}+\lbrack H_{d},f_{d}\rbrack=0,
\end{equation}
where $H_{d}=v^{2}/2+\Phi_{d}({\bf r},t)$ is the exact energy of an
individual particle and we have introduced the Poisson bracket
\begin{equation}\label{klim3}
\lbrack H,f\rbrack=\frac{\partial H}{\partial {\bf v}}\cdot \frac{\partial f}{\partial {\bf r}}-\frac{\partial H}{\partial {\bf r}}\cdot \frac{\partial f}{\partial {\bf v}}.
\end{equation}
The Klimontovich equation (\ref{klim2}) contains exactly the same
information as the Hamiltonian equations (\ref{pvm1})-(\ref{pvm2}).

We now introduce a smooth distribution function $f({\bf r},{\bf v},t)=\langle f_{d}({\bf r},{\bf v},t)\rangle$ corresponding to an average of $f_{d}({\bf r},{\bf v},t)$ over a large number of initial conditions. We then write $f_d=f+\delta f$ where $\delta f$ denotes the fluctuations around the smooth distribution. Similarly, we introduce $\Phi({\bf r},t)=\langle \Phi_{d}({\bf r},t)\rangle$ and write $\Phi_d=\Phi+\delta\Phi$. Substituting these decompositions in Eq. (\ref{klim2}), we obtain
\begin{equation}
\label{klim4}
\frac{\partial f}{\partial t}+\frac{\partial \delta f}{\partial t}+\lbrack H,f\rbrack+\lbrack H,\delta f\rbrack+\lbrack \delta\Phi, f\rbrack+\lbrack \delta\Phi, \delta f\rbrack =0,
\end{equation}
where we have written $H_{d}=H+\delta H$ with $H=v^{2}/2+\Phi$ and
$\delta H=\delta\Phi$.  Taking the ensemble average of Eq. (\ref{klim4}), we get
\begin{equation}\label{klim4b}
\frac{\partial f}{\partial t}+\lbrack H,f\rbrack=-\langle \lbrack \delta\Phi, \delta f\rbrack \rangle.
\end{equation}
Subtracting this relation from Eq. (\ref{klim4}) and neglecting the nonlinear terms, we obtain an equation for the perturbation
\begin{equation}\label{klim5}
\frac{\partial \delta f}{\partial t}+\lbrack H,\delta f\rbrack+\lbrack \delta\Phi, f\rbrack=0.
\end{equation}
This linearization corresponds to the quasilinear approximation
\cite{pitaevskii}. It is valid at the order $1/N$ in a proper
thermodynamic limit $N\rightarrow +\infty$ in which the system size scales like $V\sim 1$, the individual mass like  $m\sim 1$, the total energy like $E\sim N$, the inverse temperature like $\beta\sim 1$, the dynamical time like $t_D\sim 1$,  and the potential of interaction like $u\sim 1/N$
\cite{newpaper}. This is equivalent to putting $1/N$ in front of the potential energy in the Hamiltonian (Kac prescription). In this thermodynamic limit, the distribution function and the potential are of order $f/N\sim\Phi\sim 1$ and their fluctuations are of order $\delta f/N\sim \delta \Phi\sim 1/\sqrt{N}$ \cite{paper4}. Therefore, the term in the right hand side of Eq. (\ref{klim4b}) is of order $1/N$ (as compared to the left hand side).  For $N\rightarrow +\infty$, it vanishes
and we obtain the Vlasov equation
\begin{equation}\label{klim6}
\frac{\partial f}{\partial t}+\lbrack H,f\rbrack=0.
\end{equation}

In order to deal with spatially inhomogeneous systems, it is
convenient to use angle-action variables \cite{goldstein,btnew}. By
construction, the Hamiltonian $H$ in angle and action variables
depends only on the actions ${\bf J}=(J_1,...,J_d)$ that are constants
of the motion (the conjugate coordinates ${\bf w}=(w_1,...,w_d)$ are
called angles). Therefore, any distribution of the form $f=f({\bf J})$
is a steady state of the Vlasov equation. According to the Jeans
theorem, this is not the general form of Vlasov steady
states. However, if the potential is regular, for all practical
purposes any time-independent solution of the Vlasov equation may be
represented by a distribution of the form $f=f({\bf J})$ (strong Jeans
theorem). We assume that the system has reached a quasi stationary
state (QSS) described by a distribution of the form $f=f({\bf J})$ as
a result of a violent collisionless relaxation \cite{lb} involving
only meanfield effects. Due to finite $N$ effects (granularities,
correlations...), the distribution function $f$ will slowly evolve in
time.  This is similar to the effect of collisions in the theory of
gases. Finite $N$ effects are accounted for in the right hand side of
Eq. (\ref{klim4b}) which is similar to the collision term in the
Boltzmann equation.  This term is expected to drive the system towards
statistical equilibrium. Therefore, it is expected to select the
Boltzmann distribution among all possible steady solutions of the
Vlasov equation.  Since this term is of order $1/N$ or smaller, the
effect of ``collisions'' (granularities, finite $N$ effects,
correlations...) is a very slow process that takes place on a
timescale $Nt_{D}$ or larger. Therefore, there is a timescale
separation between the dynamical time $t_{D}\sim 1$ which is the
timescale at which the system reaches a steady state of the Vlasov
equation through violent collisionless relaxation and the
``collisional'' time $t_{R}$ at which the system is expected to relax
towards statistical equilibrium due to finite $N$ effects. Because of
this timescale separation, the distribution function will evolve
through a sequence of stationary solutions of the Vlasov equation
depending only on the actions ${\bf J}$, slowly changing with time due
to ``collisions'' (finite $N$ effects). Hence, the distribution
function averaged on a short (dynamical) timescale can be approximated
by
\begin{equation}
\label{klim7}
f({\bf r},{\bf v},t)\simeq f({\bf J},t).
\end{equation}
In that case, the distribution function is a function $f=f({\bf J},t)$
of the actions only that slowly evolves in time under the effect of
``collisions''. This is similar to an adiabatic approximation. The
system is approximately in mechanical equilibrium at each stage of the
dynamics and the ``collisions'' slowly drive it to thermodynamical
equilibrium. The purpose of this paper is to obtain a closed
expression for the collision term in action space when the rapid
dynamics has been averaged over the angles.

\subsection{Angle-action variables} \label{sec_anac}

The previous decomposition has separated the discrete distribution function $f_d$ into
a smooth component $f$ that slowly evolves with time and a fluctuating component
$\delta f$ that changes more rapidly. When we focus
on the evolution of the fluctuations, we can consider that the smooth
distribution is ``frozen''. The smooth distribution $f$
determines a smooth potential $\Phi$ that, we assume,
is integrable. Therefore, to analyze the evolution of the fluctuations,
it is possible to use  angle-action variables constructed with this smooth potential \cite{goldstein,btnew}. A particle with coordinates $({\bf r},{\bf v})$ in
phase space is described equivalently by the angle-action
variables $({\bf w},{\bf J})$. The smooth Hamiltonian equations for the
conjugate variables $({\bf r},{\bf v})$ are
\begin{equation}\label{aa8}
\frac{d{\bf r}}{dt}=\frac{\partial H}{\partial {\bf v}}, \qquad \frac{d{\bf v}}{dt}=-\frac{\partial H}{\partial {\bf r}}.
\end{equation}
If we use the variables $({\bf r},{\bf
v})$, we find that the dynamics is relatively complicated because the
potential explicitly appears in the second equation. Therefore, this
equation $d{\bf v}/dt=-\nabla\Phi$ cannot be easily integrated except
if $\Phi=0$,  i.e. for a spatially homogeneous system. In that case, the velocity ${\bf v}$ is constant
and the unperturbed equations of motion reduce to ${\bf r}={\bf v}t+{\bf
r}_{0}$, i.e. a rectilinear motion at constant velocity.  Now, the angle-action
variables are constructed so that the Hamiltonian does not depend on
the angles ${\bf w}$. Therefore, the smooth  Hamiltonian equations for the
conjugate variables  $({\bf w},{\bf J})$ are
\begin{equation}
\label{aa9}
\frac{d{\bf w}}{dt}=\frac{\partial H}{\partial {\bf J}}\equiv {\bf \Omega}({\bf J}), \qquad \frac{d{\bf J}}{dt}=-\frac{\partial H}{\partial {\bf w}}={\bf 0},
\end{equation}
where ${\bf \Omega}({\bf J})$ is
the angular frequency of the orbit with action ${\bf J}$. From these equations, we find
that ${\bf J}$ is constant and that ${\bf w}={\bf \Omega}({\bf J})t+{\bf
w}_{0}$. Therefore, the equations of
motion are very simple in these variables. They  extend naturally the
trajectories at constant velocity for spatially homogeneous systems. This is
why this choice of variables is relevant to develop the kinetic
theory. Of course, even if the description of the motion becomes simple in these variables,
the complexity of the problem has not completely disappeared. It is now embodied in the relation between position and momentum variables and angle and action variables which can be quite complicated. In this paper, we shall not consider this problem and we shall  remain in action space.

Since the Poisson bracket is invariant on a change of canonical variables, we have
\begin{equation}\label{klim3b}
\lbrack H,f\rbrack=\frac{\partial H}{\partial {\bf J}}\cdot \frac{\partial f}{\partial {\bf w}}-\frac{\partial H}{\partial {\bf w}}\cdot \frac{\partial f}{\partial {\bf J}}.
\end{equation}
Now, in terms of the angle-action variables, using the
relations
\begin{equation}\label{aa10}
\lbrack H,\delta f\rbrack=\frac{\partial H}{\partial {\bf J}}\cdot \frac{\partial \delta f}{\partial {\bf w}}-\frac{\partial H}{\partial {\bf w}}\cdot \frac{\partial \delta f}{\partial {\bf J}}={\bf \Omega}({\bf J})\cdot \frac{\partial \delta f}{\partial {\bf w}},
\end{equation}
\begin{equation}\label{aa11}
\lbrack \delta \Phi,f\rbrack=\frac{\partial \delta\Phi}{\partial {\bf J}}\cdot \frac{\partial f}{\partial {\bf w}}-\frac{\partial \delta\Phi}{\partial {\bf w}}\cdot \frac{\partial f}{\partial {\bf J}}=-\frac{\partial \delta\Phi}{\partial {\bf w}}\cdot \frac{\partial f}{\partial {\bf J}},
\end{equation}
and averaging over the angles, we find that Eqs.  (\ref{klim4b})-(\ref{klim5}) take the form
\begin{equation}
\frac{\partial f}{\partial t}=\frac{\partial}{\partial {\bf J}}\cdot \left\langle \delta f \frac{\partial\delta\Phi}{\partial {\bf w}}\right\rangle,
\label{lb3}
\end{equation}
\begin{equation}
\frac{\partial\delta f}{\partial t}+{\bf \Omega}({\bf J})\cdot \frac{\partial \delta f}{\partial {\bf w}}-\frac{\partial\delta\Phi}{\partial {\bf w}}\cdot \frac{\partial f}{\partial {\bf J}}=0.
\label{lb4}
\end{equation}
These equations govern the evolution of the smooth distribution function and of the fluctuations in angle and action space at the order $1/N$.

\subsection{Lenard-Balescu and Landau-type kinetic equations}
\label{sec_lb}

Equations (\ref{lb3}) and (\ref{lb4}) form the basis of the
quasilinear approximation. We shall assume
that the fluctuations evolve rapidly compared to the transport time
scale, so that time variation of $f({\bf J},t)$ and ${\bf \Omega}({\bf
J},t)$ can be neglected in the calculation of the collision term
(Bogoliubov ansatz). In other words, the distribution function can be
regarded as constant in Eq. (\ref{lb4}) because it evolves on a
(relaxation) timescale which is much longer than the (dynamical) time
required for the correlation function $\langle \delta
f\delta\Phi\rangle$ to reach an equilibrium for a given value of
$f$. This adiabatic hypothesis is valid for $N\gg 1$. For
brevity, we shall omit the variable $t$ in $f$ and ${\bf \Omega}$. It
will be restored at the end, when we take into account the time
dependence of the smooth distribution function through the kinetic
equation (\ref{lb3}). We also assume that the distribution function
$f$ remains Vlasov stable, so it evolves under the sole effect of
``collisions'' and not because of dynamical instabilities. In that
case, Eqs. (\ref{lb3})-(\ref{lb4}) can be solved with the aid of
Fourier-Laplace transforms and the collision term can be explicitly
calculated. We present below the calculations that lead to the
Lenard-Balescu-type equation (\ref{lb32}) for spatially inhomogeneous
systems.

The Fourier-Laplace transform of the fluctuations of the distribution
function $\delta f$ is defined by \footnote{The conventional Laplace
transform uses $-s$ in place of $-i\omega$ in the integrand. However,
we find it more convenient to use this latter notation in order to
make the link with the case where the perturbations are decomposed in
plane waves with pulsation $\omega$ and wavenumber ${\bf k}$
(e.g. when studying the Vlasov stability problem and obtaining the
dispersion relation). The notation $-i\omega$ for the Laplace transform is
relatively standard in kinetic theory.}:
\begin{equation}
\delta \tilde{f}({\bf k},{\bf J},\omega)=\int \frac{d{\bf w}}{(2\pi)^d}\int_{0}^{+\infty}dt\, e^{-i({\bf k}\cdot{\bf w}-\omega t)}\delta f({\bf w},{\bf J},t).
\label{lb5}
\end{equation}
This expression for the Laplace transform is valid for ${\rm Im}(\omega)$ sufficiently large. For the remaining part of the complex $\omega$ plane, it is defined by an analytic continuation. The inverse transform is
\begin{equation}
\delta f({\bf w},{\bf J},t)=\sum_{\bf k}\int_{\cal C}\frac{d\omega}{2\pi}\, e^{i({\bf k}\cdot{\bf w}-\omega t)}\delta {\tilde f}({\bf k},{\bf J},\omega),
\label{lb6}
\end{equation}
where the Laplace contour ${\cal C}$ in the complex $\omega$ plane must pass above all poles of the integrand. Similar expressions hold for the fluctuations of the potential $\delta\Phi({\bf w},{\bf J},t)$. If we take the  Fourier-Laplace transform of Eq. (\ref{lb4}), we find that
\begin{equation}
-\delta\hat{f}({\bf k},{\bf J},0)-i\omega\, \delta\tilde f({\bf k},{\bf J},\omega)+i {\bf k}\cdot{\bf \Omega}\, \delta\tilde f({\bf k},{\bf J},\omega)-i {\bf k}\cdot\frac{\partial f}{\partial {\bf J}}\, \delta\tilde\Phi({\bf k},{\bf J},\omega)=0,
\label{lb7}
\end{equation}
where the first term is the spatial Fourier transform of the initial value
\begin{equation}
\delta\hat{f}({\bf k},{\bf J},0)=\int\frac{d{\bf w}}{(2\pi)^d}\, e^{-i{\bf k}\cdot {\bf w}}\delta f({\bf w},{\bf J},0).
\label{lb8}
\end{equation}
The foregoing equation can be rewritten
\begin{equation}
\delta\tilde f ({\bf k},{\bf J},\omega)=\frac{{\bf k}\cdot \frac{\partial f}{\partial {\bf J}}}{{\bf k}\cdot {\bf \Omega}-\omega}\delta\tilde\Phi({\bf k},{\bf J},\omega)+\frac{\delta\hat f({\bf k},{\bf J},0)}{i({\bf k}\cdot {\bf \Omega}-\omega)},
\label{lb9}
\end{equation}
where the first term on the right hand side corresponds to ``collective effects'' and the second term is related to the initial condition. In order to simplify the notations, we have noted ${\bf\Omega}$ for ${\bf\Omega}({\bf J})$. Similarly, we shall note $f$ for $f({\bf J})$, ${\bf\Omega}'$ for ${\bf\Omega}({\bf J}')$ and  $f'$ for $f({\bf J}')$. The fluctuations of the potential $\Phi({\bf r},t)$ are related to the fluctuations of the density $\rho({\bf r},t)=\int f({\bf r},{\bf v},t)\, d{\bf v}$ by $\delta\Phi({\bf r},t)=\int u(|{\bf r}-{\bf r}'|)\delta\rho({\bf r}',t)\, d{\bf r}'$. We introduce a bi-orthogonal basis  \cite{btnew} in which the density $\rho_\alpha({\bf r})$ and the potential $\Phi_\alpha({\bf r})$ are related to each other by
\begin{equation}
\Phi_\alpha({\bf r})=\int u(|{\bf r}-{\bf r}'|)\rho_{\alpha}({\bf r}')\, d{\bf r}',\qquad \int\rho_\alpha({\bf r}) \Phi_{\alpha'}^*({\bf r})\, d{\bf r}=-\delta_{\alpha,\alpha'}.
\label{b1}
\end{equation}
The fluctuations of density $\delta\rho({\bf r},t)$ and potential  $\delta\Phi({\bf r},t)$ can be expanded on this basis as
\begin{equation}
\delta\rho({\bf r},t)=\sum_{\alpha} A_{\alpha}(t)\rho_{\alpha}({\bf r}),\qquad \delta\Phi({\bf r},t)=\sum_{\alpha} A_{\alpha}(t)\Phi_{\alpha}({\bf r}).
\label{b2}
\end{equation}
Introducing angle-action variables, we have
\begin{equation}
\delta\Phi({\bf w},{\bf J},t)=\sum_{\alpha} A_{\alpha}(t)\Phi_{\alpha}({\bf w},{\bf J}).
\label{b3}
\end{equation}
Taking the Fourier-Laplace transform of this equation, we obtain
\begin{equation}
\delta\tilde\Phi({\bf k},{\bf J},\omega)=\sum_{\alpha} \tilde{A}_{\alpha}(\omega)\hat{\Phi}_{\alpha}({\bf k},{\bf J}),
\label{b4}
\end{equation}
where
\begin{equation}
\tilde{A}_{\alpha}(\omega)=\int_{0}^{+\infty} A_{\alpha}(t) e^{i\omega t}\, dt,
\label{b5}
\end{equation}
is the Laplace transform of $A_{\alpha}(t)$ and
\begin{equation}
\hat{\Phi}_{\alpha}({\bf k},{\bf J})=\int  \Phi_{\alpha}({\bf w},{\bf J})e^{-i{\bf k}\cdot{\bf w}}\, \frac{d{\bf w}}{(2\pi)^d},
\label{b6}
\end{equation}
is the Fourier transform of $\Phi_{\alpha}({\bf w},{\bf J})$.
We can rewrite Eq. (\ref{lb9}) as
\begin{equation}
\delta\tilde f ({\bf k},{\bf J},\omega)=\frac{{\bf k}\cdot \frac{\partial f}{\partial {\bf J}}}{{\bf k}\cdot {\bf \Omega}-\omega} \sum_{\alpha} \tilde{A}_{\alpha}(\omega)\hat{\Phi}_{\alpha}({\bf k},{\bf J}) +\frac{\delta\hat f({\bf k},{\bf J},0)}{i({\bf k}\cdot {\bf \Omega}-\omega)}.
\label{b7}
\end{equation}
Taking the inverse Fourier transform of this expression, we obtain
\begin{eqnarray}
\delta\tilde{f} ({\bf w},{\bf J},\omega)=\sum_{\bf k}\, e^{i{\bf k}\cdot {\bf w}}\delta\tilde f ({\bf k},{\bf J},\omega)
=\sum_{\bf k}\, e^{i{\bf k}\cdot {\bf w}}\left\lbrack \frac{{\bf k}\cdot \frac{\partial f}{\partial {\bf J}}}{{\bf k}\cdot {\bf \Omega}-\omega} \sum_{\alpha} \tilde{A}_{\alpha}(\omega)\hat{\Phi}_{\alpha}({\bf k},{\bf J}) +\frac{\delta\hat f({\bf k},{\bf J},0)}{i({\bf k}\cdot {\bf \Omega}-\omega)}\right\rbrack.
\label{b8}
\end{eqnarray}
Multiplying the left hand side of Eq. (\ref{b8}) by $\Phi_{\alpha'}^*({\bf w},{\bf J})$ and integrating over ${\bf w}$ and ${\bf J}$, we get
\begin{eqnarray}
\int \delta\tilde{f} ({\bf w},{\bf J},\omega)\Phi_{\alpha'}^*({\bf w},{\bf J})\, d{\bf w}d{\bf J}=\int \delta\tilde{f} ({\bf r},{\bf v},\omega)\Phi_{\alpha'}^*({\bf r})\, d{\bf r}d{\bf v}
=\int \delta\tilde{\rho} ({\bf r},\omega)\Phi_{\alpha'}^*({\bf r})\, d{\bf r}\nonumber\\
=\sum_{\alpha}\int \tilde{A}_{\alpha}(\omega)\rho_{\alpha}({\bf r})\Phi_{\alpha'}^*({\bf r})\, d{\bf r}
=-\sum_{\alpha} \tilde{A}_{\alpha}(\omega)\delta_{\alpha,\alpha'}=-\tilde{A}_{\alpha'}(\omega),
\label{b9}
\end{eqnarray}
where we have used the fact that $({\bf w},{\bf J})$ are canonical variables so that $d{\bf r}d{\bf v}=d{\bf w}d{\bf J}$. Multiplying the right hand side of Eq. (\ref{b8}) by $\Phi_{\alpha'}^*({\bf w},{\bf J})$ and integrating over ${\bf w}$ and ${\bf J}$, we get
\begin{eqnarray}
\sum_{\bf k}\int d{\bf w}d{\bf J}\, e^{i{\bf k}\cdot {\bf w}}\Phi_{\alpha'}^*({\bf w},{\bf J})\left\lbrack \frac{{\bf k}\cdot \frac{\partial f}{\partial {\bf J}}}{{\bf k}\cdot {\bf \Omega}-\omega} \sum_{\alpha} \tilde{A}_{\alpha}(\omega)\hat{\Phi}_{\alpha}({\bf k},{\bf J}) +\frac{\delta\hat f({\bf k},{\bf J},0)}{i({\bf k}\cdot {\bf \Omega}-\omega)}\right\rbrack\nonumber\\
=(2\pi)^d\sum_{\bf k}\int d{\bf J}\, \hat{\Phi}_{\alpha'}^*({\bf k},{\bf J})\left\lbrack \frac{{\bf k}\cdot \frac{\partial f}{\partial {\bf J}}}{{\bf k}\cdot {\bf \Omega}-\omega} \sum_{\alpha} \tilde{A}_{\alpha}(\omega)\hat{\Phi}_{\alpha}({\bf k},{\bf J}) +\frac{\delta\hat f({\bf k},{\bf J},0)}{i({\bf k}\cdot {\bf \Omega}-\omega)}\right\rbrack.
\label{b10}
\end{eqnarray}
From Eqs. (\ref{b9}) and (\ref{b10}), we obtain
\begin{eqnarray}
-\tilde{A}_{\alpha'}(\omega)
=(2\pi)^d\sum_{\alpha}\left\lbrack \sum_{\bf k}\int d{\bf J}\, \frac{{\bf k}\cdot \frac{\partial f}{\partial {\bf J}}}{{\bf k}\cdot {\bf \Omega}-\omega} \hat{\Phi}_{\alpha}({\bf k},{\bf J})\hat{\Phi}_{\alpha'}^*({\bf k},{\bf J}) \right\rbrack  \tilde{A}_{\alpha}(\omega)
+(2\pi)^d\sum_{\bf k}\int d{\bf J}\,\frac{\delta\hat f({\bf k},{\bf J},0)}{i({\bf k}\cdot {\bf \Omega}-\omega)} \hat{\Phi}_{\alpha'}^*({\bf k},{\bf J}).
\label{b11}
\end{eqnarray}
Introducing the dielectric tensor
\begin{eqnarray}
\epsilon_{\alpha\alpha'}(\omega)=\delta_{\alpha\alpha'}+(2\pi)^d\sum_{\bf k}\int d{\bf J}\, \frac{{\bf k}\cdot \frac{\partial f}{\partial {\bf J}}}{{\bf k}\cdot {\bf \Omega}-\omega} \hat{\Phi}_{\alpha}^*({\bf k},{\bf J})\hat{\Phi}_{\alpha'}({\bf k},{\bf J}),
\label{b12}
\end{eqnarray}
the foregoing relation can be rewritten \footnote{The dispersion relation associated with the linearized Vlasov equation corresponds to ${\rm det}(\epsilon(\omega))=0$. The Vlasov dynamical stability of self-gravitating systems has been studied in \cite{kalnajsdisp,goodman88,weinbergdisp,pichoncannon,valageas1}. For the HMF model, $\epsilon(\omega)$ is a $2\times 2$ matrix. In that case, the evolution of perturbations governed by the linearized Vlasov equation has been studied by \cite{barre1,barre2} who evidenced an interesting situation of algebraic damping.}:
\begin{eqnarray}
\epsilon_{\alpha\alpha'}(\omega)\tilde{A}_{\alpha'}(\omega)=-(2\pi)^d\sum_{\bf k}\int d{\bf J}\,\frac{\delta\hat f({\bf k},{\bf J},0)}{i({\bf k}\cdot {\bf \Omega}-\omega)} \hat{\Phi}_{\alpha}^*({\bf k},{\bf J}).
\label{b13}
\end{eqnarray}
Inverting the matrix $\epsilon(\omega)$, we get
\begin{eqnarray}
\tilde{A}_{\alpha}(\omega)=-(2\pi)^d\sum_{\alpha'} (\epsilon^{-1})_{\alpha\alpha'}(\omega)\sum_{{\bf k}'}\int d{\bf J}'\,\frac{\delta\hat f({\bf k}',{\bf J}',0)}{i({\bf k}'\cdot {\bf \Omega}'-\omega)} \hat{\Phi}_{\alpha'}^*({\bf k}',{\bf J}').
\label{b14}
\end{eqnarray}
Finally, substituting Eq. (\ref{b14}) in Eq. (\ref{b4}) and introducing the notation
\begin{eqnarray}
\frac{1}{D_{{\bf k},{\bf k}'}({\bf J},{\bf J}',\omega)}=\sum_{\alpha\alpha'} \hat{\Phi}_{\alpha}({\bf k},{\bf J}) (\epsilon^{-1})_{\alpha\alpha'}(\omega) \hat{\Phi}_{\alpha'}^*({\bf k}',{\bf J}'),
\label{b16}
\end{eqnarray}
we obtain
\begin{equation}
\delta\tilde\Phi({\bf k},{\bf J},\omega)=-(2\pi)^d\sum_{{\bf k}'}\int d{\bf J}'\, \frac{1}{D_{{\bf k},{\bf k}'}({\bf J},{\bf J}',\omega)}\frac{\delta\hat f({\bf k}',{\bf J}',0)}{i({\bf k}'\cdot {\bf \Omega}'-\omega)}.
\label{b17}
\end{equation}

We can use the previous expressions to compute the collision term appearing on the right hand side of Eq. (\ref{lb3}). One has
\begin{equation}
\left\langle\delta f \frac{\partial\delta\Phi}{\partial {\bf w}}\right\rangle=\sum_{\bf k}\int_{\cal C} \frac{d\omega}{2\pi}\sum_{\bf k'}\int_{\cal C} \frac{d\omega'}{2\pi} \, i {\bf k}' e^{i({\bf k}\cdot {\bf w}-\omega t)}e^{i({\bf k}'\cdot {\bf w}-\omega' t)}\langle \delta\tilde f({\bf k},{\bf J},\omega)\delta\tilde\Phi({\bf k}',{\bf J},\omega')\rangle.
\label{lb13}
\end{equation}
Using Eq. (\ref{lb9}), we find that
\begin{equation}
\langle \delta\tilde f({\bf k},{\bf J},\omega)\delta\tilde\Phi({\bf k}',{\bf J},\omega')\rangle=\frac{{\bf k}\cdot \frac{\partial f}{\partial {\bf J}}}{{\bf k}\cdot {\bf \Omega}-\omega}\langle \delta\tilde\Phi({\bf k},{\bf J},\omega)\delta\tilde\Phi({\bf k}',{\bf J},\omega')\rangle+\frac{\langle \delta\hat f({\bf k},{\bf J},0)\delta\tilde\Phi({\bf k}',{\bf J},\omega')\rangle}{i({\bf k}\cdot {\bf \Omega}-\omega)}.
\label{lb14}
\end{equation}
The first term corresponds to the self-correlation of the potential, while the second term corresponds to the correlations between the fluctuations of the potential and of the distribution function at time $t=0$. Let us consider these two terms separately.

From Eq. (\ref{b17}), we obtain
\begin{equation}
\langle \delta\tilde\Phi({\bf k},{\bf J},\omega)\delta\tilde\Phi({\bf k}',{\bf J},\omega')\rangle=-(2\pi)^{2d}\sum_{{\bf k}_1,{\bf k}_2}\int d{\bf J}_1  d{\bf J}_2 \, \frac{1}{D_{{\bf k},{\bf k}_1}({\bf J},{\bf J}_1,\omega)} \frac{1}{D_{{\bf k}',{\bf k}_2}({\bf J},{\bf J}_2,\omega')} \frac{\langle \delta\hat f({\bf k}_1,{\bf J}_1,0)\delta\hat f({\bf k}_2,{\bf J}_2,0)\rangle}{({\bf k}_1\cdot {\bf \Omega}_1-\omega)({\bf k}_2\cdot {\bf \Omega}_2-\omega')}.
\label{lb15}
\end{equation}
Using the expression of the auto-correlation of the fluctuations at
$t=0$ given by (see Appendix \ref{sec_auto}):
\begin{equation}
\langle \delta\hat f({\bf k},{\bf J},0)\delta\hat f({\bf k}',{\bf J}',0)\rangle=\frac{1}{(2\pi)^d}\delta_{{\bf k},-{\bf k}'}\delta({\bf J}-{\bf J}') m f({\bf J}),
\label{lb16}
\end{equation}
we find that
\begin{equation}
\langle \delta\tilde\Phi({\bf k},{\bf J},\omega)\delta\tilde\Phi({\bf k}',{\bf J},\omega')\rangle=(2\pi)^{d}\sum_{{\bf k}_1}\int d{\bf J}_1  \, \frac{1}{D_{{\bf k},{\bf k}_1}({\bf J},{\bf J}_1,\omega)} \frac{1}{D_{{\bf k}',-{\bf k}_1}({\bf J},{\bf J}_1,\omega')} \frac{m f({\bf J}_1)}{({\bf k}_1\cdot {\bf \Omega}_1-\omega)({\bf k}_1\cdot {\bf \Omega}_1+\omega')}.
\label{lb17}
\end{equation}
Considering only the contributions that do not decay in time, it can be shown \cite{pitaevskii} that $\lbrack  ({\bf k}_1\cdot {\bf \Omega}_1-\omega)({\bf k}_1\cdot {\bf \Omega}_1+\omega') \rbrack^{-1}$ can be substituted by $(2\pi)^2\delta(\omega+\omega')\delta({\bf k}_1\cdot {\bf \Omega}_1-\omega)$. Therefore
\begin{equation}
\langle \delta\tilde\Phi({\bf k},{\bf J},\omega)\delta\tilde\Phi({\bf k}',{\bf J},\omega')\rangle=(2\pi)^{d+2} m \delta(\omega+\omega') \sum_{{\bf k}_1} \int d{\bf J}_1  \, \frac{1}{D_{{\bf k},{\bf k}_1}({\bf J},{\bf J}_1,\omega)} \frac{1}{D_{{\bf k}',-{\bf k}_1}({\bf J},{\bf J}_1,-\omega)}  \delta({\bf k}_1\cdot {\bf \Omega}_1-\omega)f({\bf J}_1).
\label{lb18}
\end{equation}
Similarly, one finds that the second term on the right hand side of Eq. (\ref{lb14}) is given by
\begin{equation}
\frac{\langle \delta\hat f({\bf k},{\bf J},0)\delta\tilde\Phi({\bf k}',{\bf J},\omega')\rangle}{i({\bf k}\cdot {\bf \Omega}-\omega)}=-(2\pi)^{2}m\delta(\omega+\omega')  \frac{1}{D_{{\bf k}',-{\bf k}}({\bf J},{\bf J},-\omega)} \delta({\bf k}\cdot {\bf \Omega}-\omega)f({\bf J}).
\label{lb19}
\end{equation}

From Eq. (\ref{lb18}), we get the contribution to Eq. (\ref{lb13}) of the first term of Eq. (\ref{lb14}). Since the collision term depends only on the action, it can be averaged over ${\bf w}$ without loss of information. This brings a Kronecker factor $\delta_{{\bf k},-{\bf k}'}$ which amounts to taking ${\bf k}'=-{\bf k}$. Then, using $D_{-{\bf k},-{\bf k}_1}({\bf J},{\bf J}_1,-\omega)=D_{{\bf k},{\bf k}_1}({\bf J},{\bf J}_1,\omega)^*$ \footnote{This results from Eqs. (25), (35), and (A25) of Heyvaerts \cite{heyvaerts}.},   we obtain
\begin{equation}
\left\langle\delta f \frac{\partial\delta\Phi}{\partial {\bf w}}\right\rangle_I=-i (2\pi)^{d+1} m \sum_{{\bf k},{\bf k}'}\int_{\cal C}\frac{d\omega}{2\pi} \int d{\bf J}'  \, {\bf k} \frac{{\bf k}\cdot \frac{\partial f}{\partial {\bf J}}}{{\bf k}\cdot {\bf \Omega}-\omega} \frac{1}{|D_{{\bf k},{\bf k}'}({\bf J},{\bf J}',\omega)|^2} \delta({\bf k}'\cdot {\bf \Omega}'-\omega)f({\bf J}').
\label{lb20}
\end{equation}
Finally, using the Landau prescription $\omega\rightarrow \omega+i 0^+$ and the Plemelj formula
\begin{equation}
\frac{1}{x\pm i 0^+}={\cal P}\left (\frac{1}{x}\right )\mp i\pi\delta(x),
\label{lb21}
\end{equation}
where ${\cal P}$ denotes the principal value, we can replace $1/({\bf k}\cdot {\bf\Omega}-\omega-i 0^{+})$ by $+i\pi\delta({\bf k}\cdot{\bf\Omega}-\omega)$. Then, integrating over $\omega$, we obtain
\begin{equation}
\left\langle\delta f \frac{\partial\delta\Phi}{\partial {\bf w}}\right\rangle_I=\pi (2\pi)^{d} m \sum_{{\bf k},{\bf k}'}\int d{\bf J}'  \, {\bf k}   \frac{1}{|D_{{\bf k},{\bf k}'}({\bf J},{\bf J}',{\bf k}\cdot{\bf \Omega})|^2} \delta({\bf k}\cdot {\bf \Omega}-{\bf k}'\cdot {\bf \Omega}')\left ({\bf k}\cdot \frac{\partial f}{\partial {\bf J}}\right ) f({\bf J}').
\label{lb22}
\end{equation}

From Eq. (\ref{lb19}), we get the contribution to Eq. (\ref{lb13}) of the second term of Eq. (\ref{lb14}). It is given by
\begin{equation}
\left\langle\delta f \frac{\partial\delta\Phi}{\partial {\bf w}}\right\rangle_{II}= \sum_{\bf k} \, {\bf k}\,  {\rm Im} \left (\frac{1}{D_{{\bf k},{\bf k}}({\bf J},{\bf J},{\bf k}\cdot {\bf \Omega})} \right ) m f({\bf J}).
\label{lb23}
\end{equation}
Using Eq. (\ref{b16}) and introducing the convention of summation over repeated indices, we have
\begin{equation}
\frac{1}{D_{{\bf k},{\bf k}}({\bf J},{\bf J},{\bf k}\cdot {\bf \Omega})}= \hat{\Phi}_{\alpha}({\bf k},{\bf J}) (\epsilon^{-1})_{\alpha\alpha'}({\bf k}\cdot {\bf\Omega}) \hat{\Phi}_{\alpha'}^*({\bf k},{\bf J}).
\label{lb24}
\end{equation}
Therefore
\begin{equation}
{\rm Im}\left (\frac{1}{D_{{\bf k},{\bf k}}({\bf J},{\bf J},{\bf k}\cdot {\bf \Omega})}\right )=\frac{1}{2i} \hat{\Phi}_{\alpha}({\bf k},{\bf J})\left \lbrack  (\epsilon^{-1})_{\alpha\alpha'}({\bf k}\cdot {\bf\Omega})-((\epsilon^{-1})_{\alpha'\alpha}({\bf k}\cdot {\bf\Omega}))^*\right\rbrack \hat{\Phi}_{\alpha'}^*({\bf k},{\bf J}).
\label{lb25}
\end{equation}
To compute the term in brackets, we use the identity
\begin{equation}
(\epsilon^{-1})-(\epsilon^{-1})^\dagger=\epsilon^{-1}(\epsilon^\dagger-\epsilon)(\epsilon^\dagger)^{-1}.
\label{lb26}
\end{equation}
In component form, it can be written
\begin{equation}
(\epsilon^{-1})_{\alpha\alpha'}-((\epsilon^{-1})_{\alpha'\alpha})^*=(\epsilon^{-1})_{\alpha\lambda}
(\epsilon_{\lambda'\lambda}^*-\epsilon_{\lambda\lambda'})((\epsilon^{-1})_{\alpha'\lambda'})^*.
\label{lb26b}
\end{equation}
From the expression (\ref{b12}) of  the dielectric tensor, we find that
\begin{eqnarray}
\epsilon_{\lambda'\lambda}(\omega)^*-\epsilon_{\lambda\lambda'}(\omega)=- i(2\pi)^{d+1}\sum_{\bf k}\int d{\bf J}\, \left ({\bf k}\cdot \frac{\partial f}{\partial {\bf J}}\right )\delta({\bf k}\cdot {\bf \Omega}-\omega) \hat{\Phi}_{\lambda}^*({\bf k},{\bf J})\hat{\Phi}_{\lambda'}({\bf k},{\bf J}),
\label{lb29}
\end{eqnarray}
where we have used the Landau prescription $\omega\rightarrow \omega+i 0^+$ and the Plemelj formula (\ref{lb21}). Inserting Eqs. (\ref{lb26b}) and (\ref{lb29}) in Eq. (\ref{lb25}) and using Eq. (\ref{b16}), we obtain
\begin{equation}
{\rm Im}\left (\frac{1}{D_{{\bf k},{\bf k}}({\bf J},{\bf J},{\bf k}\cdot {\bf \Omega})}\right )=-\pi (2\pi)^d\sum_{{\bf k}'}\int d{\bf J}'\,  \frac{1}{|D_{{\bf k},{\bf k}'}({\bf J},{\bf J}',{\bf k}\cdot {\bf \Omega})|^2}\delta({\bf k}\cdot {\bf \Omega}-{\bf k}'\cdot {\bf \Omega}')\left ({\bf k}'\cdot \frac{\partial f'}{\partial {\bf J}'}\right ).
\label{lb30}
\end{equation}
Substituting Eq. (\ref{lb30}) in Eq. (\ref{lb23}), we find that
\begin{equation}
\left\langle\delta f \frac{\partial\delta\Phi}{\partial {\bf w}}\right\rangle_{II}=- \pi (2\pi)^d m \sum_{{\bf k},{\bf k}'}\int d{\bf J}' \, {\bf k}\, \frac{1}{|D_{{\bf k},{\bf k}'}({\bf J},{\bf J}',{\bf k}\cdot {\bf \Omega})|^2} \delta({\bf k}\cdot {\bf \Omega}-{\bf k}'\cdot {\bf \Omega}') \left ({\bf k}'\cdot \frac{\partial f'}{\partial {\bf J}'}\right )f({\bf J}).
\label{lb31}
\end{equation}
Finally, regrouping Eqs. (\ref{lb3}), (\ref{lb22}) and (\ref{lb31}) and restoring the time variable, we end up with the kinetic equation
\begin{equation}
\frac{\partial f}{\partial t}=\pi (2\pi)^d m \frac{\partial}{\partial {\bf J}}\cdot \sum_{{\bf k},{\bf k}'}\int d{\bf J}' \, {\bf k}\, \frac{1}{|D_{{\bf k},{\bf k}'}({\bf J},{\bf J}',{\bf k}\cdot {\bf \Omega})|^2}\delta({\bf k}\cdot {\bf \Omega}-{\bf k}'\cdot {\bf \Omega}') \left ({\bf k}\cdot \frac{\partial }{\partial {\bf J}}-{\bf k}'\cdot \frac{\partial }{\partial {\bf J}'}\right )  f({\bf J},t)f({\bf J}',t).
\label{lb32}
\end{equation}

This Lenard-Balescu-type kinetic equation, taking  spatial inhomogeneity and collective effects into account, was derived by Heyvaerts  \cite{heyvaerts} from the BBGKY hierarchy. It describes the ``collisional'' evolution of the distribution function caused by the weak noise created by the discreteness of the particles accompanied by their associated polarization cloud. It is written in action space,  which is possible when the Hamiltonian associated with the average potential is integrable. For spatially homogeneous systems, it reduces to the ordinary Lenard-Balescu equation
\begin{eqnarray}
\frac{\partial f}{\partial t}=\pi (2\pi)^d m
\frac{\partial}{\partial {\bf v}}\cdot  \int\, d{\bf k}d{\bf v}'\,
{\bf k}  \frac{\hat{u}({\bf k})^2}{|\epsilon({\bf k},{\bf k}\cdot {\bf
v})|^{2}} \delta \lbrack {\bf k}\cdot ({\bf v}-{\bf v}')\rbrack {\bf k}\cdot \left (\frac{\partial}{\partial {\bf v}}-\frac{\partial}{\partial
{\bf v'}}\right ) f({\bf v},t)f({\bf v}',t)\label{lbh1}
\end{eqnarray}
where
\begin{eqnarray}
\epsilon({\bf k},\omega)=1+(2\pi)^{d}\hat{u}({\bf k})\int \frac{{\bf k}\cdot \frac{\partial f}{\partial {\bf v}}}{\omega-{\bf k}\cdot {\bf v}}\, d{\bf v}, \label{lbh2}
\end{eqnarray}
is the dielectric function. When collective effects are neglected, we obtained in Refs. \cite{angleaction,kindetail}  a Landau-type kinetic equation with angle-action variables of the form
\begin{equation}
\frac{\partial f}{\partial t}=\pi (2\pi)^d m \frac{\partial}{\partial {\bf J}}\cdot \sum_{{\bf k},{\bf k}'}\int d{\bf J}' \, {\bf k}\, |A_{{\bf k},{\bf k}'}({\bf J},{\bf J}')|^2\delta({\bf k}\cdot {\bf \Omega}-{\bf k}'\cdot {\bf \Omega}') \left ({\bf k}\cdot \frac{\partial }{\partial {\bf J}}-{\bf k}'\cdot \frac{\partial }{\partial {\bf J}'}\right )  f({\bf J},t)f({\bf J}',t),
\label{lb33}
\end{equation}
where $A_{{\bf k},{\bf k}'}({\bf J},{\bf J}')$ is the Fourier transform of the potential of interaction written in angle-action variables (see Appendix \ref{sec_pot}). This equation was derived in \cite{kindetail} by a method that does not use Fourier-Laplace transforms nor biorthogonal basis. It arises from the generalized Landau equation
\begin{eqnarray}
\frac{\partial f}{\partial t}+{\bf v}\cdot \frac{\partial f}{\partial {\bf r}}+\frac{N-1}{N} \langle {\bf
F}\rangle\cdot \frac{\partial f}{\partial {\bf v}}=\frac{\partial}{\partial {v}^{\mu}}\int_0^{t} d\tau \int d{\bf
r}_{1}d{\bf v}_1 {F}^{\mu}(1\rightarrow 0){\cal G}(t,t-\tau)\nonumber\\
\times  \left \lbrack {{\cal F}}^{\nu}(1\rightarrow 0)
{\partial\over\partial {v}^{\nu}}+{{\cal F}}^{\nu}(0\rightarrow
1) {\partial\over\partial {v}_{1}^{\nu}}\right
\rbrack f({\bf
r},{\bf v},t-\tau)\frac{f}{m}({\bf r}_1,{\bf v}_1,t-\tau), \label{lb35}
\end{eqnarray}
derived in \cite{kandrup1,paper3,paper4}, which is valid for systems
that are not necessarily spatially homogeneous and not necessarily
Markovian \footnote{The Markovian approximation may not be justified
in every situation. For example, the (exponential) decay rate of the
force auto-correlation function diverges close to a critical point
\cite{hb2} making this approximation invalid.}. The Landau-type
equation (\ref{lb33}) is obtained from Eq. (\ref{lb35}) by making a
Markovian approximation and introducing angle-action variables
\cite{kindetail}. For spatially homogeneous systems, it
reduces to the ordinary Landau equation corresponding to
Eq. (\ref{lbh1}) with $|\epsilon({\bf k},{\bf k}\cdot {\bf
v})|=1$. The connection between Eqs. (\ref{lb32}) and (\ref{lb33}) is
clear. If we neglect collective effects, ${1}/{D_{{\bf k},{\bf
k}'}({\bf J},{\bf J}',\omega)_{bare}}$ is just the Fourier transform
$A_{{\bf k},{\bf k}'}({\bf J},{\bf J}')$ of the ``bare'' potential of
interaction (see Appendix \ref{sec_pot}). When collective effects are
taken into account, the ``bare'' potential of interaction $A_{{\bf
k},{\bf k}'}({\bf J},{\bf J}')$ is replaced by the ``dressed''
potential of interaction ${1}/{D_{{\bf k},{\bf k}'}({\bf J},{\bf
J}',\omega)}$, without changing the overall structure of the kinetic
equation. This is similar to the case of homogeneous plasmas where the
bare potential ${\hat u}({\bf k})$ in the Landau equation is replaced
by the ``dressed'' potential ${\hat u}({\bf k})/|\epsilon({\bf k},{\bf
k}\cdot {\bf v})|$ in the Lenard-Balescu equation.

\subsection{The relaxation time of the system as a whole}
\label{sec_relaxwhole}

The kinetic equation (\ref{lb32}) is valid at the order $1/N$ so it describes the ``collisional'' evolution of the system on a timescale $\sim N t_D$. This kinetic equation conserves the total mass $M=\int f({\bf J},t)\, d{\bf J}$ and the total energy $E=\int f({\bf J},t)H({\bf J})\, d{\bf J}$. It also monotonically increases the Boltzmann entropy $S=-\int ({f}/{m})({\bf J},t)\ln ({f}/{m})({\bf J},t)\, d{\bf J}$ ($H$-theorem). The Boltzmann distribution $f_e({\bf J})=Ae^{-\beta m H({\bf J})}$ is a steady state of this kinetic equation. The derivation of these results can be found in \cite{angleaction,heyvaerts}. The collisional evolution of the system (at the order $1/N$) is due to a condition of resonance between distant orbits. The condition of resonance, encapsulated in the $\delta$-function, corresponds to ${\bf k}'\cdot {\bf \Omega}({\bf J}')={\bf k}\cdot {\bf \Omega}({\bf J})$ with $({\bf k}',{\bf J}')\neq ({\bf k},{\bf J})$. If there are resonances during the whole evolution, the system will relax towards the Boltzmann distribution (provided that this equilibrium state exists). In that case, the relaxation time is
\begin{equation}
t_{R}\sim N t_D.
\label{rw3}
\end{equation}
This is the case, in general, for spatially inhomogeneous systems in any dimension \footnote{For very particular situations such that ${\bf k}'\cdot {\bf \Omega}({\bf J}')\neq {\bf k}\cdot {\bf \Omega}({\bf J})$ for any $({\bf k}',{\bf J}')\neq ({\bf k},{\bf J})$ (no resonance), the relaxation time is larger than $Nt_D$.}. This is also the case for spatially homogeneous systems in $d>1$ dimensions. In that case, the condition of resonance corresponds to
${\bf k}\cdot {\bf v}'={\bf k}\cdot {\bf v}$ with ${\bf v}'\neq {\bf v}$ (the self-interaction at ${\bf v}={\bf v}'$ does not produce transport since the term in parenthesis in Eq. (\ref{lbh1}) vanishes identically).  However, for spatially homogeneous one dimensional systems, there is no resonance and the Lenard-Balescu collision term vanishes \cite{kindetail,bgm}. The kinetic equation  (\ref{lbh1}) reduces to ${\partial f}/{\partial t}=0$,
so the distribution function does not evolve at all on a timescale $\sim N t_D$. In that case, the relaxation time is larger than $Nt_D$. We therefore expect that
 \begin{equation}
t_{R}>Nt_D  \qquad ({\rm 1D  \,\, homogeneous}).
\label{rw2}
\end{equation}
Since the relaxation process is due to more complex correlations, we have to develop the kinetic theory at higher orders (taking  three-body, four-body,... correlation functions into account) to obtain the relaxation time. If the collision term does not vanish at the next order of the expansion in powers of $1/N$, the kinetic theory would imply a relaxation time of the order $N^2t_D$. However, the problem could be more complicated and yield a larger relaxation time (see the Conclusion for a more detailed discussion). In fact, it is not even granted that the system will ever relax towards statistical equilibrium. The evolution may be non-ergodic and the mixing by ``collisions'' inefficient. By contrast, for spatially homogeneous systems in $d>1$ dimensions [see Eq. (\ref{lbh1})], and for spatially inhomogeneous systems in any dimension [see Eq. (\ref{lb32})], since there are potentially more resonances, the relaxation time could be reduced and approach the natural scaling $N t_D$ predicted by the first order kinetic theory \cite{angleaction,kindetail}. However, very little is known concerning the properties of Eq. (\ref{lb32}) and its convergence (or not) towards the Boltzmann distribution. It could approach the Boltzmann distribution (since entropy increases), without reaching it exactly if the resonances disappear at some point of the evolution (see Refs. \cite{angleaction,kindetail} for more details).

\section{Stochastic process of a test particle: Diffusion and friction}
\label{sec_stoch}

\subsection{The Fokker-Planck equation}
\label{sec_fp}

In the previous section, we have studied the evolution of the system as a whole. We now consider the relaxation of a test particle is a bath of field particles with a steady distribution  $f({\bf J})$. We assume that the field particles are either at statistical equilibrium with the Boltzmann distribution (thermal bath), in which case their distribution does not change at all, or in a stable steady state of the Vlasov equation with a profile that forbids any resonance (as we have just seen, this profile does not change on a timescale of order $Nt_D$). We assume that the test particle has an initial action ${\bf J}_0$ and we study how it progressively acquires the distribution of the bath due to ``collisions'' with the field particles. As we shall see, the test particle has a stochastic motion and the evolution of the distribution function $P({\bf J},t)$, the probability density that the test particle has an action ${\bf J}$ at time $t$,  is governed by a Fokker-Planck equation involving a diffusion term and a friction term that can be analytically obtained. The Fokker-Planck equation may then be solved with the initial condition $P({\bf J},0)=\delta({\bf J}-{\bf J}_0)$ to yield $P({\bf J},t)$. This problem has been investigated in our previous papers \cite{angleaction,kindetail}, but we shall give here a direct and more rigorous derivation of the coefficients of diffusion and friction, taking collective effects into account.

The equations of motion of the test particle, written with angle-action variables, are
\begin{equation}
\label{fp1}
\frac{d{\bf w}}{dt}={\bf\Omega}({\bf J})+\frac{\partial \delta\Phi}{\partial {\bf J}}({\bf w},{\bf J},t),\qquad \frac{d{\bf J}}{dt}=-\frac{\partial\delta\Phi}{\partial {\bf w}}({\bf w},{\bf J},t).
\end{equation}
They include the effect of the mean field  which produces a zeroth order motion characterized by the pulsation ${\bf \Omega}({\bf J})$ plus a stochastic component $\delta\Phi$ of order $1/\sqrt{N}$ (fluctuations) which takes  the deviations from the mean field into account. They can be formally integrated into
\begin{equation}
\label{fp2}
{\bf w}(t)={\bf w}+\int_{0}^{t}{\bf\Omega}({\bf J}(t'))\, dt'+\int_{0}^{t}\frac{\partial\delta\Phi}{\partial {\bf J}}({\bf w}(t'),{\bf J}(t'),t')\, dt',
\end{equation}
\begin{equation}
\label{fp3}
{\bf J}(t)={\bf J}-\int_{0}^{t}\frac{\partial\delta\Phi}{\partial {\bf w}}({\bf w}(t'),{\bf J}(t'),t')\, dt',
\end{equation}
where we have assumed that, initially, the test particle is at $({\bf w},{\bf J})$. Note that the ``initial'' time considered here does not necessarily coincide with the original time mentioned above. Since the fluctuation $\delta\Phi$ of the potential is a small quantity, the foregoing equations can be solved iteratively. At the order $1/N$, which corresponds to quadratic order in $\delta\Phi$, we get for the action
\begin{eqnarray}
\label{fp4}
J_{i}(t)=J_{i}&-&\int_{0}^{t}dt'\, \frac{\partial\delta\Phi}{\partial w_{i}}({\bf w}+{\bf \Omega}t', {\bf J},t')
+\frac{\partial\Omega_{l}}{\partial J_{j}}\int_{0}^{t}dt'\int_{0}^{t'}dt''\int_{0}^{t''}dt'''\, \frac{\partial^{2}\delta\Phi}{\partial w_{i}\partial w_{l}}({\bf w}+{\bf \Omega}t', {\bf J}, t')\frac{\partial\delta\Phi}{\partial w_{j}} ({\bf w}+{\bf \Omega}t''', {\bf J}, t''')\nonumber\\
&+&\int_{0}^{t}dt'\int_{0}^{t'}dt''\, \frac{\partial^{2}\delta\Phi}{\partial w_{i}\partial J_{j}}({\bf w}+{\bf \Omega}t', {\bf J}, t')\frac{\partial\delta\Phi}{\partial w_{j}} ({\bf w}+{\bf \Omega}t'', {\bf J}, t'')\nonumber\\
&-&\int_{0}^{t}dt'\int_{0}^{t'}dt''\, \frac{\partial^{2}\delta\Phi}{\partial w_{i}\partial w_{j}}({\bf w}+{\bf \Omega}t', {\bf J}, t')\frac{\partial\delta\Phi}{\partial J_{j}} ({\bf w}+{\bf \Omega}t'', {\bf J}, t'').
\end{eqnarray}
As the changes in the action are small, the dynamics of the test particle can be represented by a stochastic process governed by a Fokker-Planck equation \cite{risken}. If we denote by $P({\bf J},t)$ the probability density that the test particle has an action ${\bf J}$ at time $t$, the general form of this equation is
\begin{equation}
\label{fp5}
\frac{\partial P}{\partial t}=\frac{\partial^{2}}{\partial J_{i}\partial J_{j}}\left (D_{ij}P\right )-\frac{\partial}{\partial J_{i}}\left (PA_{i}\right ).
\end{equation}
The diffusion tensor and the friction force are defined by
\begin{equation}
\label{fp6}
D_{ij}({\bf J})=\lim_{t\rightarrow +\infty}\frac{1}{2t} \langle (J_{i}(t)-J_{i})(J_{j}(t)-J_{j})\rangle,
\end{equation}
\begin{equation}
\label{fp7}
A_{i}({\bf J})=\lim_{t\rightarrow +\infty}\frac{1}{t} \langle J_{i}(t)-J_{i}\rangle.
\end{equation}
In writing these limits, we have implicitly assumed that the time $t$ is long with respect to the fluctuation time but short with respect to the relaxation time (of order $Nt_D$), so the expression (\ref{fp4}) can be used to evaluate Eqs. (\ref{fp6}) and (\ref{fp7}). As shown in our previous papers \cite{angleaction,kindetail}, it is relevant to rewrite the Fokker-Planck equation in the alternative form
\begin{equation}
\label{fp8}
\frac{\partial P}{\partial t}=\frac{\partial}{\partial J_{i}} \left (D_{ij}\frac{\partial P}{\partial
J_{j}}- P F_i^{pol}\right ).
\end{equation}
The total friction is
\begin{equation}
\label{fp9}
A_{i}=F_{i}^{pol}+\frac{\partial D_{ij}}{\partial J_{j}},
\end{equation}
where ${\bf F}_{pol}$ is the friction due to the polarization, while the second term is due to the variations of the diffusion tensor with ${\bf J}$. As we shall see, this decomposition arises naturally in the following analysis. The two expressions (\ref{fp5}) and (\ref{fp8}) have their own interest. The expression (\ref{fp5}) where the diffusion tensor is placed after the second derivative $\partial^2(DP)$ involves the total friction ${\bf A}$ and the expression (\ref{fp8}) where the diffusion tensor is placed between the derivatives $\partial D\partial P$ isolates the friction by polarization ${\bf F}_{pol}$.  We shall see in Sec. \ref{sec_conn} that this second form is directly related to the Lenard-Balescu-type equation (\ref{lb32}). It has therefore a clear physical interpretation. We now calculate the diffusion tensor and the friction force from Eqs. (\ref{fp6}) and (\ref{fp7}), using the results of Sec. \ref{sec_lb} that allow to take collective effects into account. The reader not interested in the details of the calculations may directly go to the final results summarized in Sec. \ref{sec_conn}.

\subsection{The diffusion tensor}
\label{sec_diff}

We first compute the diffusion tensor defined by Eq. (\ref{fp6}). Using Eq. (\ref{fp4}), we see that it is given, at the order $1/N$, by
\begin{equation}
D_{ij}=\frac{1}{2t}\int_{0}^{t}dt'\int_{0}^{t}dt'' \, \left\langle \frac{\partial\delta\Phi}{\partial w_i}({\bf w}+{\bf\Omega}t', {\bf J}, t')\frac{\partial\delta\Phi}{\partial w_j}({\bf w}+{\bf\Omega}t'', {\bf J}, t'')\right\rangle.
\label{diff1}
\end{equation}
By the inverse Fourier-Laplace transform, we have
\begin{eqnarray}
\left\langle \frac{\partial\delta\Phi}{\partial w_i}({\bf w}+{\bf\Omega}t', {\bf J}, t')\frac{\partial\delta\Phi}{\partial w_j}({\bf w}+{\bf\Omega}t'', {\bf J}, t'')\right\rangle=-\sum_{\bf k}\int_{\cal C}\frac{d\omega}{2\pi}\sum_{{\bf k}'}\int_{\cal C}\frac{d\omega'}{2\pi} \, k_i k'_j e^{i{\bf k}\cdot ({\bf w}+{\bf \Omega} t')}e^{-i\omega t'}e^{i {\bf k}'\cdot ({\bf w}+{\bf \Omega} t'')}e^{-i\omega' t''} \nonumber\\
\times\langle \delta\tilde\Phi({\bf k},{\bf J},\omega)\delta\tilde\Phi({\bf k}',{\bf J},\omega')\rangle.\qquad
\label{diff2}
\end{eqnarray}
Since the diffusion tensor depends only on the action, it can be averaged over ${\bf w}$ without loss of information. This brings a Kronecker factor
$\delta_{{\bf k},-{\bf k}'}$ which amounts to taking ${\bf k}'=-{\bf k}$. Then, substituting Eq. (\ref{lb18}) in Eq. (\ref{diff2}), and carrying out the integrals over $\omega'$ and $\omega$, we end up with the result
\begin{eqnarray}
\left\langle \frac{\partial\delta\Phi}{\partial w_i}({\bf w}+{\bf\Omega}t', {\bf J}, t')\frac{\partial\delta\Phi}{\partial w_j}({\bf w}+{\bf\Omega}t'', {\bf J}, t'')\right\rangle=(2\pi)^d m \sum_{{\bf k},{\bf k}'}\int d{\bf J}' \, k_i k_j e^{i({\bf k}\cdot {\bf \Omega}-{\bf k}'\cdot {\bf\Omega}')(t'-t'')}\frac{1}{|D_{{\bf k},{\bf k}'}({\bf J},{\bf J}',{\bf k}'\cdot {\bf\Omega}')|^2} f({\bf J}').\nonumber\\
\label{diff3}
\end{eqnarray}
This expression shows that the correlation function appearing in Eq. (\ref{diff1}) is an even function of $t'-t''$. Using the identity
\begin{eqnarray}
\int_{0}^{t}dt'\int_{0}^{t}dt''\, f(t'-t'')=2\int_{0}^{t}dt'\int_{0}^{t'}dt''\, f(t'-t'')=2\int_{0}^{t}ds\, (t-s)f(s),
\label{diff4}
\end{eqnarray}
we find, for $t\rightarrow +\infty$, that
\begin{equation}
D_{ij}=\int_{0}^{+\infty} \left\langle \frac{\partial\delta\Phi}{\partial w_i}({\bf w}, {\bf J},0)\frac{\partial\delta\Phi}{\partial w_j}({\bf w}+{\bf\Omega}s, {\bf J}, s)\right\rangle\, ds.
\label{diff5}
\end{equation}
This is the Kubo formula for our problem. Replacing the correlation function by its expression (\ref{diff3}), we get
\begin{equation}
D_{ij}=(2\pi)^d m \int_{0}^{+\infty} ds\, \sum_{{\bf k},{\bf k}'}\int d{\bf J}' \, k_i k_j e^{i({\bf k}\cdot {\bf \Omega}-{\bf k}'\cdot {\bf\Omega}')s}\frac{1}{|D_{{\bf k},{\bf k}'}({\bf J},{\bf J}',{\bf k}'\cdot {\bf\Omega}')|^2} f({\bf J}').
\label{diff6}
\end{equation}
Making the change of variables $s\rightarrow -s$ and $({\bf k},{\bf k}')\rightarrow (-{\bf k},-{\bf k}')$,  we see that we can replace $\int_{0}^{+\infty}ds$ by $(1/2)\int_{-\infty}^{+\infty}ds$. Then, using the identity
\begin{equation}
\delta(\omega)=\int_{-\infty}^{+\infty} e^{i \omega t}\, \frac{dt}{2\pi},
\label{delta}
\end{equation}
we obtain the final expression
\begin{equation}
D_{ij}=\pi(2\pi)^d m \sum_{{\bf k},{\bf k}'}\int d{\bf J}' \, k_i k_j \frac{1}{|D_{{\bf k},{\bf k}'}({\bf J},{\bf J}',{\bf k}'\cdot {\bf\Omega}')|^2} \delta({\bf k}\cdot {\bf \Omega}-{\bf k}'\cdot {\bf\Omega}') f({\bf J}').
\label{diff7}
\end{equation}

\subsection{The friction due to the polarization}
\label{sec_ess}

We now compute the friction term defined by Eq. (\ref{fp7}). We need to keep terms up to order $1/N$. From Eq. (\ref{fp4}), the first term to compute is
\begin{equation}
{\bf A}_{I}=-\frac{1}{t}\int_{0}^{t}dt'\, \left\langle \frac{\partial\delta\Phi}{\partial {\bf w}}({\bf w}+{\bf\Omega} t',{\bf J},t')\right\rangle.
\label{ess1}
\end{equation}
By the inverse Fourier-Laplace transform, we have
\begin{equation}
\left\langle \frac{\partial\delta\Phi}{\partial {\bf w}}({\bf w}+{\bf\Omega} t',{\bf J},t')\right\rangle=i\sum_{\bf k}\int_{\cal C}\frac{d\omega}{2\pi} \, {\bf k}  e^{i {\bf k}\cdot ({\bf w}+{\bf\Omega}t')}e^{-i\omega t'}\langle \delta\tilde\Phi({\bf k},{\bf J},\omega)\rangle.
\label{ess2}
\end{equation}
Using Eq. (\ref{b17}), we find that
\begin{equation}
\langle \delta\tilde\Phi({\bf k},{\bf J},\omega)\rangle=-(2\pi)^d\sum_{{\bf k}'}\int d{\bf J}'\, \frac{1}{D_{{\bf k},{\bf k}'}({\bf J},{\bf J}',\omega)}\frac{\langle \delta\hat f({\bf k}',{\bf J}',0)\rangle}{i({\bf k}'\cdot {\bf \Omega}'-\omega)}.
\label{ess3}
\end{equation}
Now, using the fact that the test particle is initially located in $({\bf w},{\bf J})$, so that $\langle \delta f({\bf w}',{\bf J}',0)\rangle=m\delta({\bf w}'-{\bf w})\delta({\bf J}'-{\bf J})$,  we obtain from Eq. (\ref{lb8}) the result
\begin{equation}
\langle \delta\hat f({\bf k}',{\bf J}',0)\rangle=\frac{1}{(2\pi)^d}m\, e^{-i{\bf k}'\cdot {\bf w}}\delta({\bf J}'-{\bf J}).
\label{ess4}
\end{equation}
Substituting these expressions in Eq. (\ref{ess2}), and averaging over ${\bf w}$ (which amounts to replacing  ${\bf k}'$ by ${\bf k}$), we get
\begin{equation}
\left\langle \frac{\partial\delta\Phi}{\partial {\bf w}}({\bf w}+{\bf\Omega} t',{\bf J},t')\right\rangle=-{m}\sum_{{\bf k}}\int_{\cal C}\frac{d\omega}{2\pi} \, {\bf k}  e^{i ({\bf k}\cdot {\bf \Omega}-\omega) t'}\frac{1}{D_{{\bf k},{\bf k}}({\bf J},{\bf J},\omega)} \frac{1}{{\bf k}\cdot {\bf \Omega}-\omega}.
\label{ess5}
\end{equation}
Therefore, the friction term (\ref{ess1}) is given by
\begin{equation}
{\bf A}_{I}=m\frac{1}{t}\int_{0}^{t}dt'\sum_{\bf k}\int_{\cal C}\frac{d\omega}{2\pi} \, {\bf k}  e^{i ({\bf k}\cdot {\bf \Omega}-\omega) t'}\frac{1}{D_{{\bf k},{\bf k}}({\bf J},{\bf J},\omega)} \frac{1}{{\bf k}\cdot {\bf \Omega}-\omega}.
\label{ess6}
\end{equation}
We now use the Landau prescription $\omega\rightarrow \omega+i 0^+$ and the Plemelj formula (\ref{lb21}) to evaluate the integral over $\omega$. The term corresponding to the imaginary part in the  Plemelj formula is
\begin{equation}
{\bf A}_{I}^{(a)}=i m \pi \frac{1}{t}\int_{0}^{t}dt'\sum_{\bf k}\int_{\cal C}\frac{d\omega}{2\pi} \, {\bf k}  e^{i ({\bf k}\cdot {\bf \Omega}-\omega) t'}\frac{1}{D_{{\bf k},{\bf k}}({\bf J},{\bf J},\omega)} \delta({\bf k}\cdot {\bf \Omega}-\omega).
\label{ess7}
\end{equation}
Integrating over $\omega$ and $t'$, we obtain
\begin{equation}
{\bf A}_{I}^{(a)}=-\frac{m}{2}\sum_{\bf k} \, {\bf k}  \, {\rm Im} \left (\frac{1}{D_{{\bf k},{\bf k}}({\bf J},{\bf J},{\bf k}\cdot {\bf\Omega})}\right ).
\label{ess8}
\end{equation}
The term corresponding to the real part in the Plemelj formula  is
\begin{equation}
{\bf A}_{I}^{(b)}=m\frac{1}{t}\int_{0}^{t}dt'\sum_{\bf k}\, {\cal P}\int_{-\infty}^{+\infty}\frac{d\omega}{2\pi} \, {\bf k}  e^{i ({\bf k}\cdot {\bf \Omega}-\omega) t'}\frac{1}{D_{{\bf k},{\bf k}}({\bf J},{\bf J},\omega)} \frac{1}{{\bf k}\cdot {\bf \Omega}-\omega}.
\label{ess10}
\end{equation}
Integrating over $t'$, we can convert this expression to the form
\begin{equation}
{\bf A}_{I}^{(b)}=-i m\sum_{\bf k}\, {\cal P}\int_{-\infty}^{+\infty}\frac{d\omega}{2\pi} \, {\bf k}\frac{1}{D_{{\bf k},{\bf k}}({\bf J},{\bf J},\omega)} \frac{1}{({\bf k}\cdot {\bf \Omega}-\omega)^2}\frac{1}{t} \left\lbrace i\sin \left\lbrack ({\bf k}\cdot {\bf\Omega}-\omega)t\right\rbrack+\cos \left\lbrack ({\bf k}\cdot {\bf\Omega}-\omega)t\right\rbrack-1\right\rbrace.
\label{ess11}
\end{equation}
For $t\rightarrow +\infty$, using the identity
\begin{equation}
\lim_{t\rightarrow +\infty}\frac{1-\cos(tx)}{t x^2}=\pi\delta(x),
\label{ess12}
\end{equation}
and integrating over $\omega$, we find that ${\bf A}_{I}^{(b)}$ is given by Eq. (\ref{ess8}), just like ${\bf A}_{I}^{(a)}$. Therefore, writing ${\bf A}_{I}={\bf A}_{I}^{(a)}+{\bf A}_{I}^{(b)}=2{\bf A}_{I}^{(a)}$, we obtain
\begin{equation}
{\bf A}_{I}=-m\sum_{\bf k} \, {\bf k}  \, {\rm Im} \left (\frac{1}{D_{{\bf k},{\bf k}}({\bf J},{\bf J},{\bf k}\cdot {\bf\Omega})}\right ).
\label{ess8b}
\end{equation}
Finally, using Eq. (\ref{lb30}), we find that
\begin{equation}
{\bf A}_{I}=\pi (2\pi)^d m\sum_{{\bf k},{\bf k}'}\int d{\bf J}'\, {\bf k} \left ({\bf k}'\cdot \frac{\partial f'}{\partial {\bf J}'}\right )\delta({\bf k}\cdot {\bf \Omega}-{\bf k}'\cdot {\bf \Omega}') \frac{1}{|D_{{\bf k},{\bf k}'}({\bf J},{\bf J}',{\bf k}\cdot {\bf \Omega})|^2}.
\label{ess14}
\end{equation}
As we shall see, ${\bf A}_{I}$ corresponds to the friction due to the polarization denoted ${\bf F}_{pol}$ in Eq. (\ref{fp9}).

{\it Remark:} We can obtain Eq. (\ref{ess14}) in a slightly more direct manner from Eq. (\ref{ess5}) by using the contour of integration shown in Fig. 9 of \cite{pitaevskii}. In that case, the integral over $\omega$ is just $-2\pi i$ times the sum of the residues at the poles of the integrand in Eq. (\ref{ess5}). The poles corresponding to the zeros of  $D_{{\bf k},{\bf k}}({\bf J},{\bf J},\omega)$ give a contribution that rapidly decays with time since ${\rm Im}(\omega)<0$ (the system is Vlasov stable). Keeping only the contribution of the pole $\omega={\bf k}\cdot {\bf \Omega}$ that does not decay in time we obtain
\begin{equation}
\left\langle \frac{\partial\delta\Phi}{\partial {\bf w}}({\bf w}+{\bf\Omega} t',{\bf J},t')\right\rangle=-i{m}\sum_{{\bf k}}\, {\bf k} \frac{1}{D_{{\bf k},{\bf k}}({\bf J},{\bf J},{\bf k}\cdot {\bf \Omega})}.
\label{ess5bis}
\end{equation}
Substituting this result in Eq. (\ref{ess1}), we get Eq. (\ref{ess8b}) then Eq. (\ref{ess14}).

\subsection{The part of the friction due to the inhomogeneity of the diffusion coefficient}
\label{sec_add}

In the evaluation of the total friction, at the order $1/N$, the second term to compute is
\begin{equation}
A_{i}^{II}=\frac{1}{t} \int_{0}^{t}dt'\int_{0}^{t'}dt''\int_{0}^{t''}dt'''\, \frac{\partial\Omega_{l}}{\partial J_{j}}\left\langle\frac{\partial^{2}\delta\Phi}{\partial w_{i}\partial w_{l}}({\bf w}+{\bf \Omega}t', {\bf J}, t')\frac{\partial\delta\Phi}{\partial w_{j}} ({\bf w}+{\bf \Omega}t''', {\bf J}, t''')\right\rangle.
\label{add1}
\end{equation}
By the inverse Fourier-Laplace transform, we have
\begin{eqnarray}
\left\langle\frac{\partial^{2}\delta\Phi}{\partial w_{i}\partial w_{l}}({\bf w}+{\bf \Omega}t', {\bf J}, t')\frac{\partial\delta\Phi}{\partial w_{j}} ({\bf w}+{\bf \Omega}t''', {\bf J}, t''')\right\rangle=-i\sum_{\bf k}\int_{\cal C}\frac{d\omega}{2\pi}\sum_{{\bf k}'}\int_{\cal C}\frac{d\omega'}{2\pi} \, k_{i}k_{l}k'_{j}   e^{i {\bf k}\cdot ({\bf w}+{\bf \Omega} t')}e^{-i\omega t'}\nonumber\\
\times e^{i {\bf k'}({\bf w}+{\bf \Omega} t''')}e^{-i\omega' t'''} \langle \delta\tilde\Phi({\bf k},{\bf J},\omega)\delta\tilde\Phi({\bf k}',{\bf J},\omega')\rangle.
\label{add2}
\end{eqnarray}
Substituting Eq. (\ref{lb18}) in Eq. (\ref{add2}), averaging over ${\bf w}$ (which amounts to replacing ${\bf k}'$ by $-{\bf k}$), and carrying out the integrals over $\omega'$ and $\omega$, we end up with the result
\begin{eqnarray}
\left\langle\frac{\partial^{2}\delta\Phi}{\partial w_{i}\partial w_{l}}({\bf w}+{\bf \Omega}t', {\bf J}, t')\frac{\partial\delta\Phi}{\partial w_{j}} ({\bf w}+{\bf \Omega}t''', {\bf J}, t''')\right\rangle=i\, (2\pi)^d m \sum_{{\bf k},{\bf k}'}\int d{\bf J}' k_i k_j k_l  e^{i ({\bf k}\cdot {\bf \Omega}-{\bf k}'\cdot {\bf\Omega}')(t'-t''')}\nonumber\\
\times \frac{1}{|D_{{\bf k},{\bf k}'}({\bf J},{\bf J}',{\bf k}'\cdot{\bf\Omega}')|^2}f({\bf J}').
\label{add3}
\end{eqnarray}
This expression shows that the correlation function appearing in Eq. (\ref{add1}) is an odd function of $t'-t'''$. Using the identity
\begin{eqnarray}
\int_{0}^{t'}dt''\int_{0}^{t''}dt'''\, f(t'-t''')=\int_{0}^{t'}dt''' \, (t'-t''')f(t'-t'''),
\label{add4}
\end{eqnarray}
we find that
\begin{equation}
A_{i}^{II}=i \, (2\pi)^d m\frac{1}{t}\int_{0}^{t}dt'\int_{0}^{t'}dt'''\sum_{{\bf k},{\bf k}'}\int d{\bf J}' \,  k_i k_j k_l \frac{\partial\Omega_{l}}{\partial J_j} (t'-t''') e^{i({\bf k}\cdot {\bf\Omega}-{\bf k}'\cdot {\bf \Omega}')(t'-t''')} \frac{1}{|D_{{\bf k},{\bf k}'}({\bf J},{\bf J}',{\bf k}'\cdot{\bf\Omega}')|^2}f({\bf J}').
\label{add5}
\end{equation}
This can be rewritten
\begin{equation}
A_{i}^{II}= (2\pi)^d m\frac{1}{t}\int_{0}^{t}dt'\int_{0}^{t'}dt'''\sum_{{\bf k},{\bf k}'}\int d{\bf J}'\,  k_i k_j \frac{\partial}{\partial J_j}\left (e^{i({\bf k}\cdot {\bf\Omega}-{\bf k}'\cdot {\bf \Omega}')(t'-t''')}\right ) \frac{1}{|D_{{\bf k},{\bf k}'}({\bf J},{\bf J}',{\bf k}'\cdot{\bf\Omega}')|^2}f({\bf J}'),
\label{add6}
\end{equation}
or, equivalently,
\begin{eqnarray}
A_{i}^{II}=(2\pi)^d m\frac{1}{t}\frac{\partial}{\partial J_{j}}\int_{0}^{t}dt'\int_{0}^{t'}dt'''\sum_{{\bf k},{\bf k}'}\int d{\bf J}' \,  k_i k_j e^{i({\bf k}\cdot {\bf\Omega}-{\bf k}'\cdot {\bf \Omega}')(t'-t''')} \frac{1}{|D_{{\bf k},{\bf k}_{1}}({\bf J},{\bf J}',{\bf k}'\cdot{\bf\Omega}')|^2}f({\bf J}')\nonumber\\
-(2\pi)^d m\frac{1}{t}\int_{0}^{t}dt'\int_{0}^{t'}dt'''\sum_{{\bf k},{\bf k}'}\int d{\bf J}' \,  k_i k_j e^{i({\bf k}\cdot {\bf\Omega}-{\bf k}'\cdot {\bf \Omega}')(t'-t''')} \frac{\partial}{\partial J_j}\left (\frac{1}{|D_{{\bf k},{\bf k}'}({\bf J},{\bf J}',{\bf k}'\cdot{\bf\Omega}')|^2}\right )f({\bf J}').
\label{add7}
\end{eqnarray}
Since the integrand only depends on $t'-t'''$, using the identity (\ref{diff4}), we obtain for $t\rightarrow +\infty$,
\begin{eqnarray}
A_{i}^{II}=(2\pi)^d m\frac{\partial}{\partial J_{j}}\int_{0}^{+\infty}ds\sum_{{\bf k},{\bf k}'}\int d{\bf J}' \,  k_i k_j e^{i({\bf k}\cdot {\bf\Omega}-{\bf k}'\cdot {\bf \Omega}')s} \frac{1}{|D_{{\bf k},{\bf k}'}({\bf J},{\bf J}',{\bf k}'\cdot{\bf\Omega}')|^2}f({\bf J}')\nonumber\\
-(2\pi)^d m\int_{0}^{+\infty}ds\sum_{{\bf k},{\bf k}'}\int d{\bf J}'\,  k_i k_j e^{i({\bf k}\cdot {\bf\Omega}-{\bf k}'\cdot {\bf \Omega}')s} \frac{\partial}{\partial J_j}\left (\frac{1}{|D_{{\bf k},{\bf k}'}({\bf J},{\bf J}',{\bf k}'\cdot{\bf\Omega}')|^2}\right )f({\bf J}').
\label{add8}
\end{eqnarray}
Making the change of variables $s\rightarrow -s$ and $({\bf k},{\bf k}')\rightarrow (-{\bf k},-{\bf k}')$,  we see that we can replace $\int_{0}^{+\infty}ds$ by $(1/2)\int_{-\infty}^{+\infty}ds$. Then, using the identity (\ref{delta}), we obtain the expression
\begin{eqnarray}
A_{i}^{II}=\pi (2\pi)^d m\frac{\partial}{\partial J_{j}}\sum_{{\bf k},{\bf k}'}\int d{\bf J}' \,  k_i k_j \frac{1}{|D_{{\bf k},{\bf k}'}({\bf J},{\bf J}',{\bf k}'\cdot{\bf\Omega}')|^2}\delta({\bf k}\cdot {\bf\Omega}-{\bf k}'\cdot {\bf \Omega}')f({\bf J}')\nonumber\\
-\pi (2\pi)^d m\sum_{{\bf k},{\bf k}'}\int d{\bf J}' \,  k_i k_j  \frac{\partial}{\partial J_j}\left (\frac{1}{|D_{{\bf k},{\bf k}'}({\bf J},{\bf J}',{\bf k}'\cdot{\bf\Omega}')|^2}\right )\delta({\bf k}\cdot {\bf\Omega}-{\bf k}'\cdot {\bf \Omega}') f({\bf J}').
\label{add9a}
\end{eqnarray}
In the first term, we recover the diffusion coefficient (\ref{diff7}) so that
\begin{eqnarray}
A_{i}^{II}=\frac{\partial D_{ij}}{\partial J_{j}}
-\pi (2\pi)^d m\sum_{{\bf k},{\bf k}'}\int d{\bf J}' \,  k_i k_j \delta({\bf k}\cdot {\bf\Omega}-{\bf k}'\cdot {\bf \Omega}') \frac{\partial}{\partial J_j}\left (\frac{1}{|D_{{\bf k},{\bf k}'}({\bf J},{\bf J}',{\bf k}'\cdot{\bf\Omega}')|^2}\right )f({\bf J}').
\label{add9}
\end{eqnarray}

Finally, the third and fourth terms to compute are
\begin{equation}
A_{i}^{III}=\frac{1}{t}\int_{0}^{t}dt'\int_{0}^{t'}dt''\, \left\langle\frac{\partial^{2}\delta\Phi}{\partial w_{i}\partial J_{j}}({\bf w}+{\bf \Omega}t', {\bf J}, t')\frac{\partial\delta\Phi}{\partial w_{j}} ({\bf w}+{\bf \Omega}t'', {\bf J}, t'')  \right\rangle,
\label{add13}
\end{equation}
and
\begin{equation}
A_{i}^{IV}=-\frac{1}{t}\int_{0}^{t}dt'\int_{0}^{t'}dt''\, \left\langle\frac{\partial^{2}\delta\Phi}{\partial w_{i}\partial w_{j}}({\bf w}+{\bf \Omega}t', {\bf J}, t')\frac{\partial\delta\Phi}{\partial J_{j}} ({\bf w}+{\bf \Omega}t'', {\bf J}, t'')\right\rangle.
\label{add14}
\end{equation}
Substituting the inverse Fourier-Laplace transform of the fluctuations of the potential in these equations, averaging over ${\bf w}$ (which amounts to replacing ${\bf k}'$ by $-{\bf k}$) and summing the resulting expressions, we obtain
\begin{equation}
A_i^{III}+A_i^{IV}=\frac{1}{t}\int_{0}^{t}dt'\int_{0}^{t'}dt''\sum_{\bf k}\int_{\cal C}\frac{d\omega}{2\pi}\int_{\cal C}\frac{d\omega'}{2\pi} \, k_i k_j e^{i {\bf k}\cdot {\bf\Omega}t'}e^{-i\omega t'}e^{-i {\bf k}\cdot {\bf\Omega}t''}e^{-i\omega' t''} \frac{\partial}{\partial J_j}\langle \delta\tilde\Phi({\bf k},{\bf J},\omega)\delta\tilde\Phi(-{\bf k},{\bf J},\omega')\rangle.
\label{add15}
\end{equation}
Substituting Eq. (\ref{lb18}) in Eq. (\ref{add15}), and carrying out the integrals over $\omega'$ and $\omega$, we end up with the result
\begin{eqnarray}
A_i^{III}+A_i^{IV}=(2\pi)^d m \frac{1}{t}\int_{0}^{t}dt'\int_{0}^{t'}dt''\sum_{{\bf k},{\bf k}'}\int d{\bf J}' \, k_i k_j e^{i ({\bf k}\cdot {\bf\Omega}-{\bf k}'\cdot {\bf\Omega}')(t'-t'')} \frac{\partial}{\partial J_j}\left (\frac{1}{|D_{{\bf k},{\bf k}'}({\bf J},{\bf J}',{\bf k}'\cdot{\Omega}')|^2}\right )f({\bf J}').
\label{add16}
\end{eqnarray}
This expression shows that the correlation function appearing under the integral sign only depends on $t'-t''$. Using the identity (\ref{diff4}) we find, for $t\rightarrow +\infty$, that
\begin{eqnarray}
A_i^{III}+A_i^{IV}=(2\pi)^d m \int_{0}^{+\infty}ds\sum_{{\bf k},{\bf k}'}\int d{\bf J}' \, k_i k_j e^{i ({\bf k}\cdot {\bf\Omega}-{\bf k}'\cdot {\bf\Omega}')s} \frac{\partial}{\partial J_j}\left (\frac{1}{|D_{{\bf k},{\bf k}'}({\bf J},{\bf J}',{\bf k}'\cdot{\Omega}')|^2}\right )f({\bf J}').
\label{add17}
\end{eqnarray}
Making the change of variables $s\rightarrow -s$ and $({\bf k},{\bf k}')\rightarrow (-{\bf k},-{\bf k}')$,  we see that we can replace $\int_{0}^{+\infty}ds$ by $(1/2)\int_{-\infty}^{+\infty}ds$. Then, using the identity (\ref{delta}), we obtain the expression
\begin{equation}
A_i^{III}+A_i^{IV}=\pi (2\pi)^d m \sum_{{\bf k},{\bf k}'}\int d{\bf J}' \, k_i k_j  \frac{\partial}{\partial J_j}\left (\frac{1}{|D_{{\bf k},{\bf k}'}({\bf J},{\bf J}',{\bf k}'\cdot{\Omega}')|^2}\right )\delta({\bf k}\cdot {\bf\Omega}-{\bf k}'\cdot {\bf\Omega}') f({\bf J}').
\label{add18}
\end{equation}
Finally, summing Eqs. (\ref{add9}) and (\ref{add18}), we get
\begin{equation}
A_{II}+A_{III}+A_{IV}=\frac{\partial D_{ij}}{\partial J_{j}}.
\label{add19}
\end{equation}

\subsection{Connection between the kinetic equation (\ref{lb32}) and the Fokker-Planck equation (\ref{fp8})}
\label{sec_conn}

We have established that the diffusion tensor and the friction force are given by
\begin{equation}
D_{ij}=\pi(2\pi)^d m \sum_{{\bf k},{\bf k}'}\int d{\bf J}'\, k_i k_j \frac{1}{|D_{{\bf k},{\bf k}'}({\bf J},{\bf J}',{\bf k}'\cdot {\bf\Omega}')|^2} \delta({\bf k}\cdot {\bf \Omega}-{\bf k}'\cdot {\bf\Omega}') f({\bf J}').
\label{diff7again}
\end{equation}
\begin{equation}
A_i=\pi (2\pi)^d m\sum_{{\bf k},{\bf k}'}\int d{\bf J}'\,  {k}_i  \frac{1}{|D_{{\bf k},{\bf k}'}({\bf J},{\bf J}',{\bf k}\cdot {\bf \Omega})|^2}\delta({\bf k}\cdot {\bf \Omega}-{\bf k}'\cdot {\bf \Omega}')\left ({\bf k}'\cdot \frac{\partial f'}{\partial {\bf J}'}\right )+\frac{\partial D_{ij}}{\partial J_{j}}.
\label{add20again}
\end{equation}
Comparing Eq. (\ref{add20again}) with Eq. (\ref{fp9}), we see that the friction by polarization is
\begin{equation}
{\bf F}_{pol}=\pi (2\pi)^d m\sum_{{\bf k},{\bf k}'}\int d{\bf J}'\, {\bf k} \frac{1}{|D_{{\bf k},{\bf k}'}({\bf J},{\bf J}',{\bf k}\cdot {\bf \Omega})|^2}\delta({\bf k}\cdot {\bf \Omega}-{\bf k}'\cdot {\bf \Omega}') \left ({\bf k}'\cdot \frac{\partial f'}{\partial {\bf J}'}\right ).
\label{ess14again}
\end{equation}
On the other hand, using an integration by part in the first term of Eq. (\ref{add20again}), the total friction can be written
\begin{equation}
{\bf A}=\pi (2\pi)^d m\sum_{{\bf k},{\bf k}'}\int d{\bf J}'\, f({\bf J}') \left ({\bf k}\cdot \frac{\partial}{\partial {\bf J}}-{\bf k}'\cdot \frac{\partial}{\partial {\bf J}'}\right )   \frac{1}{|D_{{\bf k},{\bf k}'}({\bf J},{\bf J}',{\bf k}\cdot {\bf \Omega})|^2}\delta({\bf k}\cdot {\bf \Omega}-{\bf k}'\cdot {\bf \Omega}').
\label{conn4}
\end{equation}
Finally, using Eqs. (\ref{diff7again}) and (\ref{ess14again}), we find that the Fokker-Planck equation (\ref{fp8}) becomes
\begin{equation}
\frac{\partial P}{\partial t}=\pi (2\pi)^d m \frac{\partial}{\partial {\bf J}}\cdot \sum_{{\bf k},{\bf k}'}\int d{\bf J}' \, {\bf k}\, \frac{1}{|D_{{\bf k},{\bf k}'}({\bf J},{\bf J}',{\bf k}\cdot {\bf \Omega})|^2}\delta({\bf k}\cdot {\bf \Omega}-{\bf k}'\cdot {\bf \Omega}') \left ({\bf k}\cdot \frac{\partial }{\partial {\bf J}}-{\bf k}'\cdot \frac{\partial }{\partial {\bf J}'}\right )  P({\bf J},t)f({\bf J}').
\label{conn5}
\end{equation}
When collective effects are neglected, we recover the results obtained in \cite{angleaction,kindetail} by a different method (related expressions of the diffusion and friction terms have also been obtained in \cite{lbk,tw,bl,wei,nt,w,pa,valageas2}). For spatially homogeneous systems, we recover the results of Hubbard \cite{hubbard} (see also \cite{paper4} when collective effects are neglected). As observed in our previous works, we note that the form of Eq. (\ref{conn5}) is very similar to the form of Eq. (\ref{lb32}). This shows that the Fokker-Planck equation (\ref{conn5}), with the diffusion coefficient (\ref{diff7again}) and the friction (\ref{ess14again}), can be directly obtained from the Lenard-Balescu-type  equation (\ref{lb32}) by replacing the time dependent distribution $f({\bf J}',t)$ by the {\it static} distribution $f({\bf J}')$ of the bath. This procedure transforms an integro-differential equation (\ref{lb32}) into a  differential equation (\ref{conn5}).  Although natural, the rigorous justification of this procedure requires the detailed calculation that we have given here. In fact, we can understand this result in the following manner. Equations (\ref{lb32}) and (\ref{conn5}) govern the evolution of the distribution function of a test particle (described by the coordinate ${\bf J}$) interacting with field particles (described by the running coordinate ${\bf J}'$). In Eq. (\ref{lb32}), all the particles are equivalent so the distribution of the field particles $f({\bf J}',t)$ changes with time exactly like the distribution of the test particle $f({\bf J},t)$. In Eq. (\ref{conn5}), the test particle and the field particles are not equivalent since the field particles form a ``bath''. The field particles have a steady (given) distribution $f({\bf J}')$ while the distribution of the test particle $f({\bf J},t)=N m P({\bf J},t)$ changes with time.

\subsection{Thermal bath: Boltzmann distribution}
\label{sec_tb}

For a thermal bath, the field particles have the Boltzmann distribution of statistical equilibrium
\begin{eqnarray}
f({\bf J})=A e^{-\beta m H({\bf J})}, \label{tb1}
\end{eqnarray}
where $H({\bf J})$ is the individual energy of the particles. We note the identity
\begin{eqnarray}
\frac{\partial f}{\partial {\bf J}}=-\beta m f({\bf J}){\bf\Omega}({\bf J}),
\label{tb2}
\end{eqnarray}
where we have used $\partial H/\partial {\bf J}={\bf\Omega}({\bf J})$. Substituting this relation in Eq. (\ref{ess14again}), we obtain
\begin{eqnarray}
{\bf F}_{pol}=-\pi (2\pi)^d m^2\beta\sum_{{\bf k},{\bf k}'}\int d{\bf J}'\,  {\bf k} \frac{1}{|D_{{\bf k},{\bf k}'}({\bf J},{\bf J}',{\bf k}\cdot {\bf \Omega})|^2}\delta({\bf k}\cdot {\bf \Omega}-{\bf k}'\cdot {\bf \Omega}') ({\bf k}'\cdot {\bf\Omega}')  f({\bf J}').
\label{tb3}
\end{eqnarray}
Using the $\delta$-function to replace ${\bf k}'\cdot {\bf\Omega}'$ by ${\bf k}\cdot {\bf\Omega}$,  and comparing the resulting expression with Eq. (\ref{diff7again}), we find that
\begin{eqnarray}
F_i^{pol}=-\beta m D_{ij}({\bf J})\Omega_j({\bf J}). \label{tb4}
\end{eqnarray}
This can be viewed as a generalized Einstein relation connecting the friction force to the diffusion tensor (fluctuation-dissipation theorem). We stress that the Einstein relation is valid for the friction by polarization  ${\bf F}_{pol}$, not for the total friction  ${\bf A}$ that has a more complex expression due to the term $\partial_j D_{ij}$.  We do not have this subtlety for the usual Brownian motion where the diffusion coefficient is constant. For a thermal bath, using
Eq. (\ref{tb4}), the Fokker-Planck equation (\ref{fp8}) takes the
form
\begin{equation}
\label{tb6}\frac{\partial P}{\partial t}=\frac{\partial}{\partial J_{i}} \left \lbrack D_{ij}({\bf J})\left (\frac{\partial P}{\partial
J_{j}}+\beta m  P \Omega_j({\bf J})\right )\right\rbrack,
\end{equation}
where $D_{ij}({\bf J})$ is given by Eq. (\ref{diff7again}) with
Eq. (\ref{tb1}). Recalling that ${\bf\Omega}({\bf
J})=\partial H/\partial {\bf J}$, this equation is similar to
the Kramers equation in Brownian theory \cite{kramers}. This is a
drift-diffusion equation describing the evolution of the distribution
$P({\bf J},t)$ of the test particle in an ``effective potential''
$U_{eff}({\bf J})=H({\bf J})$ produced by the field
particles. For $t\rightarrow +\infty$, the distribution of the test
particle relaxes towards the Boltzmann distribution $P_e({\bf
J})=(A/Nm)e^{-\beta m H({\bf J})}$. Since the Fokker-Planck is valid at the
order $1/N$, the
relaxation time scales like
\begin{equation}
t_{R}^{bath}\sim {N}t_D.
\label{rbath1}
\end{equation}

If the field particles are in a stable steady state of the Vlasov equation that forbids any resonance (see Sect. \ref{sec_relaxwhole}), the diffusion tensor and the friction force can be simplified in a manner similar to that described in \cite{angleaction,kindetail}. In that case, the test particle still relaxes towards the distribution $f({\bf J})$ of the bath on a timescale $t_{R}^{bath}\sim {N}t_D$. This timescale is shorter than the relaxation time of the system as a whole $t_R>Nt_D$ so that the bath approximation, which assumes that the distribution of the field particles is ``frozen'',  is justified.

\section{Conclusion}
\label{sec_conclusion}

In this paper, we have developed a kinetic theory of systems with
long-range interactions which takes collective effects and spatial
inhomogeneity into account. This improves our previous works
\cite{angleaction,kindetail} where collective effects were neglected
(leading to the Landau-type kinetic equation (\ref{lb33})) or taken
into account in an unsatisfactory manner (making an {\it ad hoc}
factorization assumption). The use of a biorthogonal basis of pairs of
density-potential functions solves the shortcomings of our previous
derivation. We recovered and confirmed the Lenard-Balescu-type kinetic
equation (\ref{lb32}) derived by Heyvaerts \cite{heyvaerts} from the
Liouville equation, using the BBGKY hierarchy truncated at the level
of the two-body distribution function (i.e. neglecting three-body
correlations). Our derivation, starting from the Klimontovich equation
and using a quasilinear approximation, may be technically simpler
\footnote{Of course, it is well-known that the approach based on the
first two equations of the BBGKY hierarchy (neglecting three-body
correlations) is equivalent to the approach starting from the
Klimontovich equation and making a quasilinear approximation. More
generally, the whole BBGKY hierarchy can be reconstructed from the
Klimontovich equation
\cite{fried,ichimaru}. However, it has been acknowledged in plasma
physics that the Klimontovich approach provides a technically simpler
and more transparent derivation of the Lenard-Balescu equation than
the one starting from the BBGKY hierarchy
\cite{klimontovich,fried}.}. Furthermore, it is completely parallel to
the one exposed in
\cite{pitaevskii,cdr,newpaper} to derive the ordinary Lenard-Balescu
equation for spatially homogeneous systems and in \cite{dubin,pvnew}
to derive a Lenard-Balescu-type equation appropriate to
two-dimensional point vortices with axisymmetric distribution. Our
approach therefore offers a unified description of kinetic theories
for systems with long-range interactions. Interestingly, the same
formalism can be used to describe the relaxation of a test particle in
a bath of field particles through a Fokker-Planck equation. This is
another advantage of our approach. We have obtained the expressions of
the diffusion tensor and friction force directly from the equations of
motion with angle-action variables, taking collective effects into
account. We have shown that they can also be obtained from the
Lenard-Balescu-type kinetic equation (\ref{lb32}) by making a bath
approximation leading to the Fokker-Planck equation (\ref{conn5}). For
simplicity of notation, we have considered a single species system,
but the general case of a multi-species system can be treated easily
and leads to the same results as those obtained by Heyvaerts
\cite{heyvaerts}.

The kinetic theory developed in this paper is valid at the order $1/N$
so it describes the evolution of the system on a timescale $Nt_D$. The
dynamical evolution of the system is due to a condition of resonance
between distant orbits. For spatially homogeneous one dimensional
systems, there is no resonance so the Lenard-Balescu collision term
vanishes (this result is well-known in plasma physics \cite{feix,kp}
and it has been rediscovered recently for the HMF model
\cite{bd,cvb}). This implies that the relaxation time is larger than
$Nt_D$. In that case, we need to develop the kinetic theory at higher
orders (taking three-body, four-body,... correlations into
account). At present, no such theory exists and the scaling of the
relaxation time with $N$ remains an open problem. We can try to make
speculations based on available results of numerical simulations. The
most natural scaling would be $N^2 t_D$ which corresponds to the next
order term in the expansion of the basic equations of the kinetic
theory in powers of $1/N$ \cite{kindetail}. An $N^2$ scaling is indeed
observed numerically \cite{dawson,rouetfeix} for spatially homogeneous
one-dimensional plasmas (in that context, $N$ represents the number of
charges in the Debye sphere). However, the scaling of the relaxation
time may be more complex. For example, for the permanently spatially
homogeneous HMF model, Campa {\it et al.} \cite{campa} report a
relaxation time scaling like $e^N t_D$. This timescale is, however,
questioned in recent works \cite{private} who find a $N^2t_D$
scaling. This would be more natural on a theoretical point of view.
It could also happen that the system never reaches statistical
equilibrium, i.e. the evolution may be non-ergodic and the mixing by
``collisions'' inefficient. This may be the case for the
two-dimensional point vortex gas when the profile of angular velocity
is monotonic \cite{dubin,pre,pvnew}. In that case, there is no
resonance and the relaxation time is larger than $Nt_D$. However, the
precise scaling with $N$ is still an open problem, and even the
relaxation towards the Boltzmann distribution is uncertain. The
situation is different for spatially homogeneous systems in $d>1$
dimensions and for spatially inhomogeneous systems in any dimension
(and for non-axisymmetric configurations of the point vortex gas). In
these cases, there are potentially more resonances so the relaxation
time can be reduced and achieve the natural scaling $Nt_D$
corresponding to the first order of the kinetic theory
\cite{kindetail}. An $N t_D$ scaling is indeed observed numerically
for spatially inhomogeneous one dimensional stellar systems
\cite{brucemiller,gouda,valageas2,joyce} and for the spatially
inhomogeneous HMF model \cite{ruffoN}. On the other hand, for the HMF
model, if an initially spatially homogeneous distribution function
becomes Vlasov unstable during the collisional evolution, a dynamical
phase transition from a non-magnetized to a magnetized state takes
place (as theoretically studied in \cite{campachav}) and the
relaxation time could be {\it intermediate} between $N^2 t_D$
(permanently homogeneous) and $N t_D$ (permanently inhomogeneous). In
that situation, Yamaguchi {\it et al.} \cite{yamaguchi} find a
relaxation time scaling like $N^{\delta}t_D$ with $\delta=1.7$. The
previous argument (suggesting $1<\delta<2$) may provide a first step
towards the explanation of this anomalous exponent.

The results of this paper also apply to self-gravitating systems that are, by essence, spatially inhomogeneous and limited in extension. The consideration of spatial inhomogeneity avoids the logarithmic divergence occurring at large scales when one makes the local approximation \cite{btnew}. There remains, however, a logarithmic divergence at small scales due to the neglect of strong collisions in the kinetic theory. On the other hand, self-gravitating systems are very particular since no statistical equilibrium state exists in a strict sense (the entropy is not bounded from above) \cite{paddy,katz,ijmpb}. Therefore, the Lenard-Balescu-type kinetic equation (\ref{lb32}) does not relax towards a steady state. On the contrary, the system takes a ``core-halo'' structure and keeps evolving. The halo expands due to the evaporation of high energy stars \cite{spitzerevap} and, when the system becomes sufficiently centrally condensed,  the core collapses as a result of the gravothermal catastrophe \cite{antonov,lbw} caused by the negative specific heat of the central region. This evolution is clearly illustrated in the numerical simulation of Cohn \cite{cohn} based on the orbit-averaged Fokker-Planck equation. Core collapse ultimately leads to the formation of binary stars \cite{henon}, followed by a post-collapse evolution \cite{il},  and gravothermal oscillations \cite{sb}. The formation of binary stars is not taken into account in the kinetic theory because it results from triple collisions (three-body correlations) and is of the strong interaction type. We also note that, since the Boltzmann distribution is not normalizable for self-gravitating systems (infinite mass problem), the thermal bath approximation that we have developed in Sec. \ref{sec_tb} is not justified. However, it could be easily generalized by replacing the Boltzmann distribution by the King model which takes evaporation into account and leads to a truncated Boltzmann distribution with a finite mass viewed as a metastable equilibrium state \cite{ijmpb}.

At the physical level, one interest of the kinetic theory is to show
that the ``collisional'' evolution of the system is controlled by
resonances encapsulated in a $\delta$-function. This explains the
scaling of the relaxation time with $N$. For example, in the case of
one dimensional potentials, the relaxation time of spatially
homogeneous systems is larger than $Nt_D$ because there is no
resonance, while it is of order $Nt_D$ for spatially inhomogeneous
systems due to the arising of resonances. Obtaining this result was
our original motivation to develop a kinetic theory for spatially
inhomogeneous systems in \cite{angleaction,kindetail}. At a practical
level, we believe that the Lenard-Balescu equation (\ref{lb32}) with
angle-action variables is very complicated to solve, so its practical
interest is limited. With present-day computers, it is much easier to
solve the $N$-body problem directly. Although the description of the
motion becomes simple in angle-action variables, all the complexity is
now embodied in the relation between position-momentum variables and
angle-action variables. The usefulness of equation (\ref{lb32}) is
therefore limited to systems for which this relation can be
established, either analytically or numerically. Some simplification
can be gained if we restrict ourselves to spherically symmetric
stellar systems as done by Heyvaerts \cite{heyvaerts} for stellar
systems. But, even in that case the final kinetic equations remain
complicated and approximations may be welcome to simplify the problem
and lead to more tractable equations. The Fokker-Planck treatment that
we have given here is one possible approximation. The restriction to
one dimensional systems with simple potentials (e.g. the HMF model) is
another one.

\appendix

\section{Bare and dressed potentials of interaction}
\label{sec_pot}

We assume that the particles interact via a long-range binary potential $u(|{\bf r}-{\bf r}'|)$ so that
\begin{eqnarray}
\Phi({\bf r},t)=\int u(|{\bf r}-{\bf r}'|) f({\bf r}',{\bf v}',t)\, d{\bf r}'d{\bf v}'.
\label{pot1}
\end{eqnarray}
We introduce angle-action variables and decompose the potential of interaction in Fourier modes
\begin{eqnarray}
u(|{\bf r}({\bf w},{\bf J})-{\bf r}'({\bf w}',{\bf J}')|)=\sum_{{\bf k},{\bf k}'}e^{i({\bf k}\cdot {\bf w}-{\bf k}'\cdot{\bf w}')}A_{{\bf k},{\bf k}'}({\bf J},{\bf J}'),
\label{pot2}
\end{eqnarray}
where the Fourier coefficients are given by
\begin{eqnarray}
A_{{\bf k},{\bf k}'}({\bf J},{\bf J}')=\frac{1}{(2\pi)^{2d}}\int u(|{\bf r}({\bf w},{\bf J})-{\bf r}'({\bf w'},{\bf J}')|)e^{-i({\bf k}\cdot {\bf w}-{\bf k}'\cdot {\bf w}')}\, d{\bf w}d{\bf w}'.
\label{pot2b}
\end{eqnarray}
The fluctuations of the potential are related to the fluctuations of
the distribution function by
\begin{eqnarray}
\delta\Phi({\bf w},{\bf J},t)=\int u(|{\bf r}-{\bf r}'|) \delta f({\bf w}',{\bf J}',t)\, d{\bf w}'d{\bf J}',
\label{pot3}
\end{eqnarray}
where we have used the identity  $d{\bf r}'d{\bf v}'=d{\bf w}'d{\bf J}'$ valid for a canonical change of variables. Taking the Fourier-Laplace transform of this expression, and using Eqs. (\ref{lb6}) and (\ref{pot2b}), we get
\begin{eqnarray}
\delta\tilde\Phi({\bf k},{\bf J},\omega)=(2\pi)^d\sum_{{\bf k}'}\int d{\bf J}' A_{{\bf k},{\bf k}'}({\bf J},{\bf J}') \delta \tilde f({\bf k}',{\bf J}',\omega).
\label{pot4}
\end{eqnarray} 
Substituting Eq. (\ref{lb9}) in Eq. (\ref{pot4}), we obtain the Fredholm integral equation
\begin{eqnarray}
\delta\tilde\Phi({\bf k},{\bf J},\omega)-(2\pi)^d\sum_{{\bf k}'}\int d{\bf J}' A_{{\bf k},{\bf k}'}({\bf J},{\bf J}') \frac{{\bf k}'\cdot \frac{\partial f'}{\partial {\bf J}'}}{{\bf k}'\cdot {\bf \Omega}'-\omega}\delta\tilde\Phi({\bf k}',{\bf J}',\omega)=(2\pi)^d\sum_{{\bf k}'}\int d{\bf J}' A_{{\bf k},{\bf k}'}({\bf J},{\bf J}')\frac{\delta\hat f({\bf k}',{\bf J}',0)}{i({\bf k}'\cdot {\bf \Omega}'-\omega)}.
\label{pot5}
\end{eqnarray}
This equation relates the Fourier-Laplace transform of the
fluctuations of the potential to the Fourier transform of the
initial fluctuations of the distribution function. It is
therefore equivalent to Eq. (\ref{b17}). When collective effects are
neglected, the foregoing expression reduces to
\begin{eqnarray}
\delta\tilde\Phi({\bf k},{\bf J},\omega)=(2\pi)^d\sum_{{\bf k}'}\int d{\bf J}' A_{{\bf k},{\bf k}'}({\bf J},{\bf J}')\frac{\delta\hat f({\bf k}',{\bf J}',0)}{i({\bf k}'\cdot {\bf \Omega}'-\omega)}.
\label{pot6}
\end{eqnarray}
Comparing this expression with Eq. (\ref{b17}), we see that $-{1}/{D_{{\bf k},{\bf k}'}({\bf J},{\bf J}',\omega)_{bare}}$ is just the Fourier transform $A_{{\bf k},{\bf k}'}({\bf J},{\bf J}')$ of the ``bare'' potential of interaction $u$ with angle-action variables. When collective effects are taken into account, Eq. (\ref{pot6}) is replaced by  Eq. (\ref{b17}) where according to (\ref{pot5})  the ``dressed'' potential of interaction satisfies
\begin{eqnarray}
\frac{1}{D_{{\bf k},{\bf k}'}({\bf J},{\bf J}',\omega)}-(2\pi)^d\sum_{{\bf k}''}\int d{\bf J}'' A_{{\bf k},{\bf k}''}({\bf J},{\bf J}'') \frac{{\bf k}''\cdot \frac{\partial f''}{\partial {\bf J}''}}{{\bf k}''\cdot {\bf \Omega}''-\omega}\frac{1}{D_{{\bf k}'',{\bf k}'}({\bf J}'',{\bf J}',\omega)}=-A_{{\bf k},{\bf k}'}({\bf J},{\bf J}').
\label{pot7}
\end{eqnarray}

\section{The resolvant}
\label{sec_resolvant}

Substituting Eq. (\ref{b17})  in Eq. (\ref{lb9}), we obtain
\begin{equation}
\delta\tilde f ({\bf k},{\bf J},\omega)=-\frac{{\bf k}\cdot \frac{\partial f}{\partial {\bf J}}}{{\bf k}\cdot {\bf \Omega}-\omega}(2\pi)^d\sum_{{\bf k}''}\int d{\bf J}''\, \frac{1}{D_{{\bf k},{\bf k}''}({\bf J},{\bf J}'',\omega)}\frac{\delta\hat f({\bf k}'',{\bf J}'',0)}{i({\bf k}''\cdot {\bf \Omega}''-\omega)}+\frac{\delta\hat f({\bf k},{\bf J},0)}{i({\bf k}\cdot {\bf \Omega}-\omega)}.
\label{res1}
\end{equation}
This equation relates the Fourier-Laplace transform of the
fluctuations of the distribution function to the Fourier
transform of the initial fluctuations of the distribution function. It can be written
\begin{equation}
\delta\tilde f ({\bf k},{\bf J},\omega)=\sum_{{\bf k}''}\int d{\bf J}''\, R_{{\bf k},{\bf k}''}({\bf J},{\bf J}'',\omega)\delta\hat{f}({\bf k}'',{\bf J}'',0),
\label{res2}
\end{equation}
where
\begin{equation}
R_{{\bf k},{\bf k}''}({\bf J},{\bf J}'',\omega)=-\frac{{\bf k}\cdot \frac{\partial f}{\partial {\bf J}}}{{\bf k}\cdot {\bf \Omega}-\omega}(2\pi)^d \frac{1}{D_{{\bf k},{\bf k}''}({\bf J},{\bf J}'',\omega)}\frac{1}{i({\bf k}''\cdot {\bf \Omega}''-\omega)}+\frac{\delta({\bf J}-{\bf J}'')}{i({\bf k}\cdot {\bf \Omega}-\omega)}\delta_{{\bf k},{\bf k}''},
\label{res3}
\end{equation}
is called the resolvant. If we consider an initial condition of the form $\delta f({\bf w},{\bf J},0)=m\delta({\bf w}-{\bf w}')\delta({\bf J}-{\bf J}')$ implying
\begin{equation}
\delta\hat{f}({\bf k},{\bf J},0)=\frac{m}{(2\pi)^d}e^{-i{\bf k}\cdot {\bf w}'}\delta({\bf J}-{\bf J}'),
\label{res4}
\end{equation}
we find that
\begin{equation}
\delta\tilde f ({\bf k},{\bf J},\omega)=-\frac{{\bf k}\cdot \frac{\partial f}{\partial {\bf J}}}{{\bf k}\cdot {\bf \Omega}-\omega}\sum_{{\bf k}''}\, 
m e^{-i{\bf k}''\cdot {\bf w}'}\frac{1}{D_{{\bf k},{\bf k}''}({\bf J},{\bf J}',\omega)}\frac{1}{i({\bf k}''\cdot {\bf \Omega}'-\omega)}+\frac{m}{(2\pi)^d}e^{-i{\bf k}\cdot {\bf w}'}\delta({\bf J}-{\bf J}')\frac{1}{i({\bf k}\cdot {\bf \Omega}-\omega)}.
\label{res5}
\end{equation}
This quantity is called the propagator of the linearized Vlasov equation.
If we substitute Eq. (\ref{pot4}) in Eq. (\ref{lb9}), we obtain the
integral equation
\begin{equation}
\delta\tilde f ({\bf k},{\bf J},\omega)=\frac{{\bf k}\cdot \frac{\partial f}{\partial {\bf J}}}{{\bf k}\cdot {\bf \Omega}-\omega}(2\pi)^d\sum_{{\bf k}'}\int d{\bf J}' A_{{\bf k},{\bf k}'}({\bf J},{\bf J}') \delta \tilde f({\bf k}',{\bf J}',\omega)+\frac{\delta\hat f({\bf k},{\bf J},0)}{i({\bf k}\cdot {\bf \Omega}-\omega)}.
\label{res6}
\end{equation}
This equation relates the Fourier-Laplace transform of the
fluctuations of the distribution function to the Fourier transform of
the initial fluctuations of the distribution function, so it is
equivalent to Eq. (\ref{res1}).

\section{Auto-correlation of the fluctuations of the one-particle distribution}
\label{sec_auto}

According to Eq. (\ref{lb8}), we have
\begin{eqnarray}
\langle \delta\hat f({\bf k},{\bf J},0)\delta\hat f({\bf k}',{\bf J}',0)\rangle&=&\int\frac{d{\bf w}}{(2\pi)^d}\frac{d{\bf w}'}{(2\pi)^d}e^{-i({\bf k}\cdot {\bf w}+{\bf k}'\cdot {\bf w}')}\langle \delta f({\bf w},{\bf J},0)\delta f({\bf w}',{\bf J}',0)\rangle\nonumber\\
&=&\int\frac{d{\bf w}}{(2\pi)^d}\frac{d{\bf w}'}{(2\pi)^d}e^{-i({\bf k}\cdot {\bf w}+{\bf k}'\cdot {\bf w}')}\left\lbrack \langle f_d({\bf w},{\bf J},0)f_d({\bf w}',{\bf J},0)\rangle-f({\bf J})f({\bf J}')\right\rbrack.
\label{auto1}
\end{eqnarray}
The expression of the discrete distribution function recalled in the first paragraph of Sec. \ref{sec_klim} leads to
\begin{eqnarray}
\langle f_d({\bf w},{\bf J},0)f_d({\bf w}',{\bf J}',0)\rangle&=&m^2\sum_{i,j}\left\langle \delta({\bf w}-{\bf w}_i)\delta({\bf J}-{\bf J}_i)\delta({\bf w}'-{\bf w}_j)\delta({\bf J}'-{\bf J}_j)\right\rangle\nonumber\\
&=&m^2\sum_{i}\left\langle \delta({\bf w}-{\bf w}_i)\delta({\bf J}-{\bf J}_i)\delta({\bf w}-{\bf w}')\delta({\bf J}-{\bf J}')\right\rangle\nonumber\\
&+&m^2\sum_{i\neq j}\left\langle \delta({\bf w}-{\bf w}_i)\delta({\bf J}-{\bf J}_i)\delta({\bf w}'-{\bf w}_j)\delta({\bf J}'-{\bf J}_j)\right\rangle\nonumber\\
&=& m f({\bf J}) \delta({\bf w}-{\bf w}')\delta({\bf J}-{\bf J}')+f({\bf J})f({\bf J}'),
\label{auto2}
\end{eqnarray}
where we have assumed that there is no correlation initially (if there are initial correlations, it can be shown that they are washed out rapidly \cite{pitaevskii} so they have no effect on the final form of the collision term). Combining Eqs. (\ref{auto1}) and (\ref{auto2}), we obtain
\begin{equation}
\langle \delta\hat f({\bf k},{\bf J},0)\delta\hat f({\bf k}',{\bf J}',0)\rangle=\frac{1}{(2\pi)^d}\delta_{{\bf k},-{\bf k}'}\delta({\bf J}-{\bf J}') m f({\bf J}).
\label{auto3}
\end{equation}

\end{document}